\documentclass[prb,aps,twocolumn,superscriptaddress,showpacs,footinbib,longbibliography]{revtex4-2}
\usepackage[colorlinks=true,linkcolor=black, citecolor=blue, urlcolor=blue, 
unicode=true]{hyperref}

\usepackage{hyperref}
\usepackage{graphicx}
\usepackage{epstopdf}
\usepackage{xcolor}
\usepackage{amsmath}

\begin{document}

\title{Magnetoelastic honeycomb fragmentation in VI$_{3}$}
\author{Enlin Shen}
\affiliation{School of Physics and Astronomy, University of Edinburgh, EH9 3JZ, United Kingdom}
\author{Tiberiu I. Popescu}
\affiliation{School of Physics and Astronomy, University of Edinburgh, EH9 3JZ, United Kingdom}
\author{Nishwal Gora}
\affiliation{School of Physics and Astronomy, University of Edinburgh, EH9 3JZ, United Kingdom}
\author{Guratinder Kaur}
\affiliation{School of Physics and Astronomy, University of Edinburgh, EH9 3JZ, United Kingdom}
\author{Edmond Chan}
\affiliation{School of Physics and Astronomy, University of Edinburgh, EH9 3JZ, United Kingdom}
\author{Harry Lane}
\affiliation{School of Physics and Astronomy, University of Edinburgh, EH9 3JZ, United Kingdom}
\author{Jose A. Rodriguez-Rivera}
\affiliation{NIST Center for Neutron Research, 100 Bureau Drive, Gaithersburg, Maryland 20899, USA}
\affiliation{Department of Materials Science, University of Maryland, College Park, MD  20742}
\author{Guangyong Xu}
\affiliation{NIST Center for Neutron Research, 100 Bureau Drive, Gaithersburg, Maryland 20899, USA}
\author{Peter M. Gehring}
\affiliation{NIST Center for Neutron Research, 100 Bureau Drive, Gaithersburg, Maryland 20899, USA}
\author{Russell A. Ewings}
\affiliation{ISIS Facility, Rutherford Appleton Laboratory, Didcot, OX11 0QX, U.K.}
\author{Andy N. Fitch}
\affiliation{European Synchrotron Radiation Facility, BP 220, 38043, Grenoble Cedex, France}
\author{Chris Stock}
\affiliation{School of Physics and Astronomy, University of Edinburgh, EH9 3JZ, United Kingdom}

\date{\today}

\begin{abstract}

The unexpected discovery of ordered magnetism in two-dimensional van der Waals materials at the monolayer limit [B. Huang \textit{et al.} \href{https://www.nature.com/articles/nature22391}{Nature {\bf{546}}, 270 (2017)}] challenges the Mermin-Wagner theorem [N. D. Mermin and H. Wagner \href{https://doi.org/10.1103/PhysRevLett.17.1133}{Phys. Rev. Lett. {\bf{17}}, 133 (1966)}], which forbids spontaneous breaking of continuous symmetries in two dimensions at finite temperatures. The persistence of static magnetism in low-dimensions is fundamentally influenced by magnetic anisotropy which is tied to the local single-ion crystalline electric field. Crucially, spin-orbit coupling connects the structural properties with spin degrees of freedom. We investigate the magnetic single-ion properties in the two-dimensional van der Waals magnet VI$_3$. Utilizing neutron and x-ray diffraction, we map out the symmetry breaking phase transitions in VI$_{3}$ and argue for the presence of a single structural transition at T$_S \sim$ 80 K, driven by an orbital degeneracy, followed by a ferromagnetic transition at a lower temperature, T$_C \sim$ 50 K. Through a comparative analysis of samples prepared under varying conditions, we suggest that lower temperature transitions reported near $\sim$ 30 K are not intrinsic to VI$_{3}$. A group theoretical analysis suggests a structural transition from rhombohedral $R\overline{3}$ to triclinic $P\overline{1}$ or $P1$. This transition is significant as it suggests the formation of two distinct crystallographyically inequivalent V$^{3+}$ sites on the honeycomb lattice, each with distinct spin-orbital properties. Neutron spectroscopy provides evidence for dominant magnetic exchange coupling only between symmetry-equivalent sites in the triclinc unit cell. We suggest this breaks up the low-temperature two-dimensional honeycomb VI$_3$ lattice into two interpenetrating approximately hexagonal planes resulting in a fragmentated honeycomb. Our findings highlight the critical role of magnetoelastic coupling in determining the magnetic and structural phases in two-dimensional van der Waals magnets.

\end{abstract}

\maketitle

\section{Introduction}

Magnetism in two dimensions~\cite{Burch18:563} is marginal with it being forbidden in isotropic ferromagnets~\cite{Mermin66:17,Mermin68:176,Hohenberg67:158}, and only viable in the presence of local single-ion anisotropy.  Indeed, even truly two dimensional structures are borderline with x-ray diffraction experiments in two dimensional liquid crystals~\cite{Als-Nielsen80:22} providing an illustration of algebraic decay of density correlations.  Anisotropy is central to ordering in magnets with reduced dimensions and even in providing a barrier to disorder as illustrated by experimental~\cite{Hill91:66,Hill97:55} and theoretical~\cite{Imry75:35} studies on random fields in model magnets.  The spin-orbit interaction plays a central role in determining local magnetic anisotropy~\cite{Yosida:book}, hence in this work we investigate the structural and orbital physics in a two dimensional magnet which is expected to host an orbital degeneracy and therefore spin-orbit coupling.  We discuss the structural and magnetic phase transitions in the context of single-ion physics in two-dimensional VI$_{3}$.

\begin{figure*}
	\includegraphics[width=170mm]{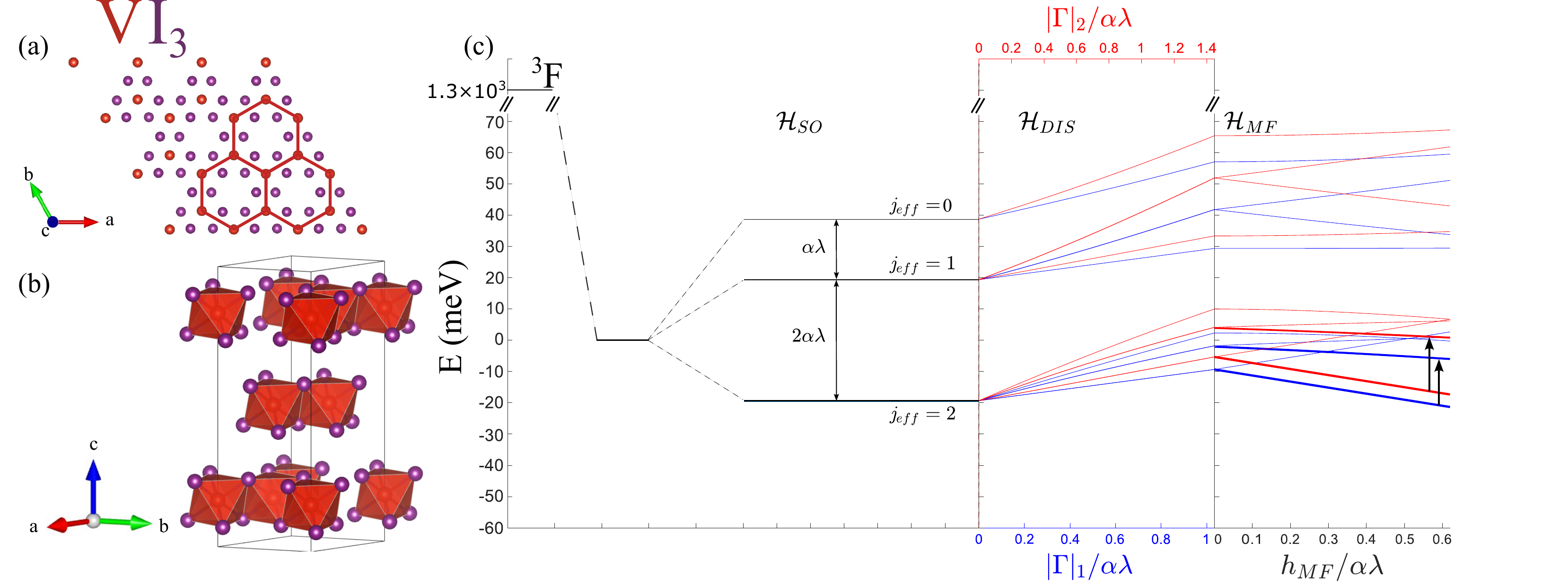}
	\caption{The $(a)$, crystal structure of VI$_3$ in the $a-b$ plane together with the $(b)$, unit cell of VI$_3$ emphasizing the stacking of the honeycomb sheets. A plot of the $(c)$ single-ion splitting due to spin-orbit coupling, tetragonal distortion and the molecular field is also presented.  The lowest energy dipole allowed transitions are indicated by the arrow.}
	\label{fig:E-level splitting}
\end{figure*}

VI$_{3}$~\cite{Yang20:101,Kong19:31} is a two dimensional van der Waals magnet which is based on a planar honeycomb lattice with threefold coordination of V$^{3+}$ ions within the $a-b$ plane as illustrated in Fig. \ref{fig:E-level splitting} $(a)$ and $(b)$.  The local octahedral environment results in the 2-$d$ electrons occupying the low-energy $|t_{2g}\rangle$ orbital manifold.  Applying Hunds rules, this results in a net spin $S=1$ and an orbital degeneracy with an effective orbital angular momentum of $l_{eff}=1$.  At high temperatures, the V$^{3+}$ ions are stacked in an ideal $ABC$ arrangement along $c$, in a structure with space group $R\overline{3}$.  Given the orbital degeneracy and following the Jahn-Teller theorem~\cite{Gehring75:38} for orbital degenerate systems, a structural transition occurs at $T_{S}\sim 80$ K.  A lower temperature transition to ferromagnetic order is reported at T$_{C} \sim 50$ K.~\cite{Kong19:31,Tian19}

The effect of spin-orbital properties on the quantum magnetism of the V$^{3+}$ sites is illustrated in Fig. \ref{fig:E-level splitting} $(c)$ with spin-orbit coupling characterized by the parameter $\alpha \lambda$.  The local octahedral crystalline electric field environment combined with a Hund's rule imposed $|S=1,l_{eff}=1\rangle$ ground state results in three spin-orbit split magnetic states defined by differing total $j_{eff}$ angular momentum imposed by the addition theorem of angular momentum and a ground state with $j_{eff}$=2.  Structural distortions (characterized by the parameter $\Gamma$ and discussed in the main text) of this local environment subtly change the ground state splitting into a series of non- Kramers doublets with the twofold degeneracy only broken by a time reversal symmetry breaking Zeeman field imposed by low temperature magnetic order described by the molecular field $h_{MF}$ in Fig. \ref{fig:E-level splitting} $(c)$.  These single-ion states provide the groundwork for our ``excitonic" theory of the magnetism in VI$_{3}$ discussed in this paper and previously~\cite{Lane21:104_2} that we use to describe neutron spectroscopy data.   

There have been a number of varied reports discussing the series of structural and magnetic symmetry breaking transitions in VI$_{3}$~\cite{Son19:99} which exist to monolayer limit~\cite{Lin21:21}.  While nearly all studies report a high temperature structural transition at T$_{S}\sim 80$ K and a lower temperature ferromagnetic transition at T$_{C}\sim$ 50 K~\cite{Liu20:1}, there has not been consensus on the low temperature structural symmetry.  Several lower temperature space groups have been reported and additional structural transitions at T $\sim$ 30 K have been measured.~\cite{Dolezal19:3,Marchandier21:104}  

Applying symmetry constraints imposed by group theory, we argue for the presence of a single structural transition at T$_{S}\sim 80$ K followed by a lower temperature magnetic transition at T$_{C} \sim 50$ K.  This results in two distinct magnetic sites with differing spin-orbital ground states, which we probe with spectroscopy.  We speculate and propose that a lower temperature transition may occur that depends on the stacking of the two-dimensional VI$_{3}$ layers.  To understand the impact of the magnetostructural transitions on V$^{3+}$ single-ion properties, we perform x-ray diffraction on a series of samples and re-examine existing neutron spectroscopy data.  This paper is divided into five parts including this introduction.  We discuss the experiments used for both probing the structural and magnetic transitions along with a description of our materials preparation.  We then discuss powder x-ray diffraction taken on crushed single crystals and three different powders made under differing heating conditions.  We then discuss the consequences of the phase transition on the magnetism by revisiting neutron spectroscopy data.  We finish the paper with conclusions and discussion.

\section{Experiment}

Single crystals of VI$_{3}$ were made using the vapor transport technique.~\cite{Juza}   Sealed quartz ampoules with outer diameter 18 mm and inner diameter of 16 mm were loaded with 3 g of vanadium and iodine in stoichiometric quantities. Approximately 5\% excess of iodine, by mass, was included to act as a transport agent. The vanadium powder was initially pumped to less than 10$^{-5}$ Torr using a turbo pump to ensure dryness before the iodine was loaded. Iodine pellets were then added and the combined reagents were chilled (to avoid iodine sublimation) and pumped to 5 $\times$ 10$^{-3}$ Torr using an oil based mechanical pump to avoid potential damage to the blades of the turbo pump from iodine vapor. The tubes were sealed to be a length of 15 cm and put into a 3-zone furnace such that one end was at 400$^{\circ}$C and the other end was at 350$^{\circ}$C. A chiller was used to further cool one end of the furnace to increase the temperature gradient. The temperature gradient was initially inverted for 12 hours to clean one end of the ampoules. 12 ampoules were loaded for each crystal growth run that lasted $\sim$ 10 days. The quartz ampoules were then removed at high temperature with one end immediately cooled using compressed air and water on removal from the three-zone furnace. A variety of crystal sizes resulted in dimensions up to a maximum of 5 mm $\times$ 5 mm, with a width of less than 1 mm.  Some of these crystals were lightly ground in a nitrogen filled dry box (with humidity less than 2 \%) for powder x-ray diffraction which we discuss below.  We note that extensive grinding of the single crystals resulted in a complete loss of measurable Bragg peaks illustrating total destruction of the sample.  We discuss the structural and magnetic transitions in these single crystals below applying neutron diffraction and review previously published spectroscopy data.  All VI$_{3}$ samples were found to be very air-sensitive~\cite{Kratochvilova22:278} degrading completely after approximately 1 hour exposure to air.  For neutron spectrosopy measurements, the use of Fomblin oil was found to significantly reduce this degradation so that coated samples did not observably degrade on the timescale of several weeks to one month, particularly when stored with water absorbing desiccant.

Powder samples of VI$_{3}$ were made by sealing vanadium and iodine in a quartz ampoule with excess iodine.  To remove iodine contamination, after high temperature treatment, from the samples, we used a sublimation technique, sealing the samples in a sublimation device under nitrogen atmosphere in a dry box (less than 2\% humidity) and cooling the cold head with ice while heating the base to $\sim$ 50$^{\circ}$C to remove iodine.  Three different powder samples were made using a box furnace, with increased temperature homogeneity preventing single crystal growth in an attempt to minimize preferred orientation of resulting samples.  The first sample (labelled Sample A for the purposes of this paper) was made by heating stoichiometric amounts of vanadium and iodine in a box furnace at 400 $^\circ$C for 3 days followed by quenching.  A second sample (Sample B) was made at 500 $^\circ$C and gradually cooled to room temperature over a period of $\sim$ 12 hours.  A third sample (Sample C) was made at 400 $^\circ$C and gradually cooled to room temperature.  We discuss the structural phase transitions of these three different samples.

We further studied the structural and magnetic transitions in our single crystals using neutron diffraction carried out on the SPINS cold triple-axis spectrometer and the MACS cold triple-axis spectrometer (NIST, USA).~\cite{Rodriguez08:19} In all experiments on single crystals the samples were mounted such that Bragg reflections of the form (HHL) lay within the horizontal scattering plane.  On SPINS, a vertically focused PG(002) monochromator was used to select an incident energy of E$_{i}$=5.0 meV. A PG(002) analyzer was used to fix the final energy to the elastic scattering condition with E$_{f}$=5.0 meV. Cooled Beryllium filters were placed both in the incident and scattering beams to reduce higher order contamination. The horizontal beam collimation sequence was set to $guide$-80$'$-$Sample$-80$'$-$open$.  Further high-intensity diffraction allowing separation of the magnetic transitions was performed on MACS with a double focused PG(002) monochromator used to select an incident energy and focus the neutron beam onto the sample. 20 PG(002) double bounce analyzers were then used to select a final energy of E$_{f}$=3.5 meV. Cooled Beryllium filters were used on the scattered side of the sample to remove higher order contamination from the sample.  In this paper we compare calculations against published neutron spectroscopy results~\cite{Lane21:104_2} taken on the MACS triple-axis spectrometer (NIST, USA) and the MAPS chopper spectrometer (ISIS, UK) to understand the implications of the structural transition on the single-ion magnetism.

X-ray powder diffraction was performed using a Rigaku Smartlab equiped with a Cu source and a wavelength of $\lambda=$1.54 \AA\ fixed by a Johansson monochromator.  Measurements were done in reflection Bragg-Bretano geometry utilizing parabolic Cross Beam Optics (CBO).~\cite{Osakabe17:33}  Samples were loaded on a plate in a dry box and then quickly moved into the vacuum environment of a PheniX displex (for access to temperatures as low as 15 K), being exposed to air for less than several minutes.   Further high resolution x-ray diffraction measurements were performed at the European Synchrotron Radiation Facility (ESRF) on the instrument ID22 using a wavelength of $\lambda$ =0.40 \AA.  Measurements on ID22 were performed in transmission mode using a glass capillary loaded in an Argon glovebox.

\section{Diffraction and Structure}

In this section we outline the crystal structure and phase transitions in our VI$_{3}$ samples.  We evaluate four different samples including single crystals grown by vapor transport and the three powder samples made under the different conditions discussed above.  

\begin{figure}
	\centering
	\includegraphics[width=75mm,trim=1.3cm 2.7cm 0.75cm 1.2cm,clip=true]{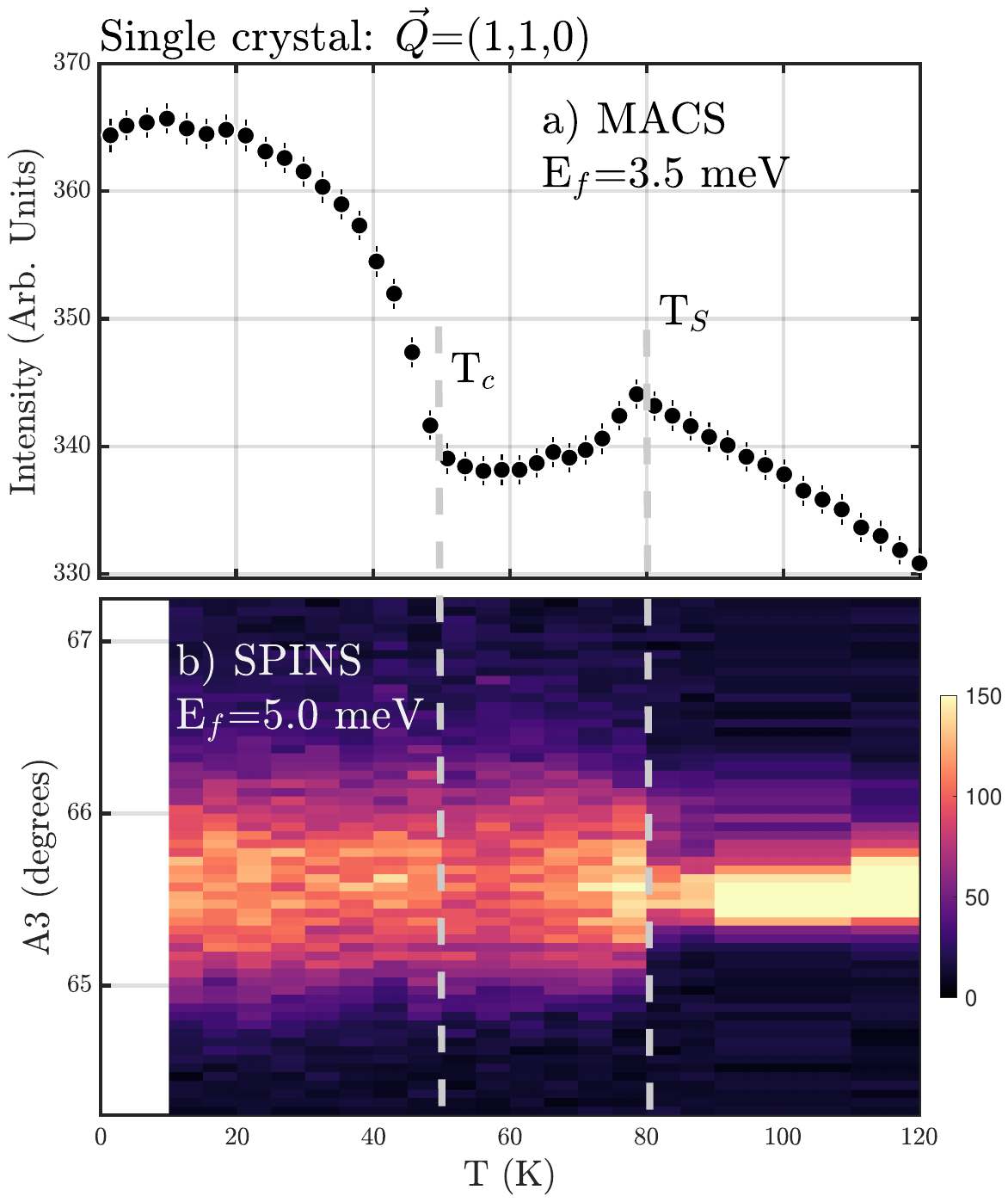}
	\caption{The temperature dependence of the the $\vec{Q}$=(1,1,0) Bragg peak taken on $(a)$ MACS and $(b)$ SPINS with scans in A3.  The ferromagnetic transition T$_{c}$ and the structural transition are denoted by dashed lines. }
	\label{fig:neutron_elastic}
\end{figure}

\subsection{Single crystal samples}
\begin{figure*}
	\centering
	\begin{tabular}{cc}
		\parbox{0.55\textwidth}{
			\centering
			\includegraphics[width=\linewidth]{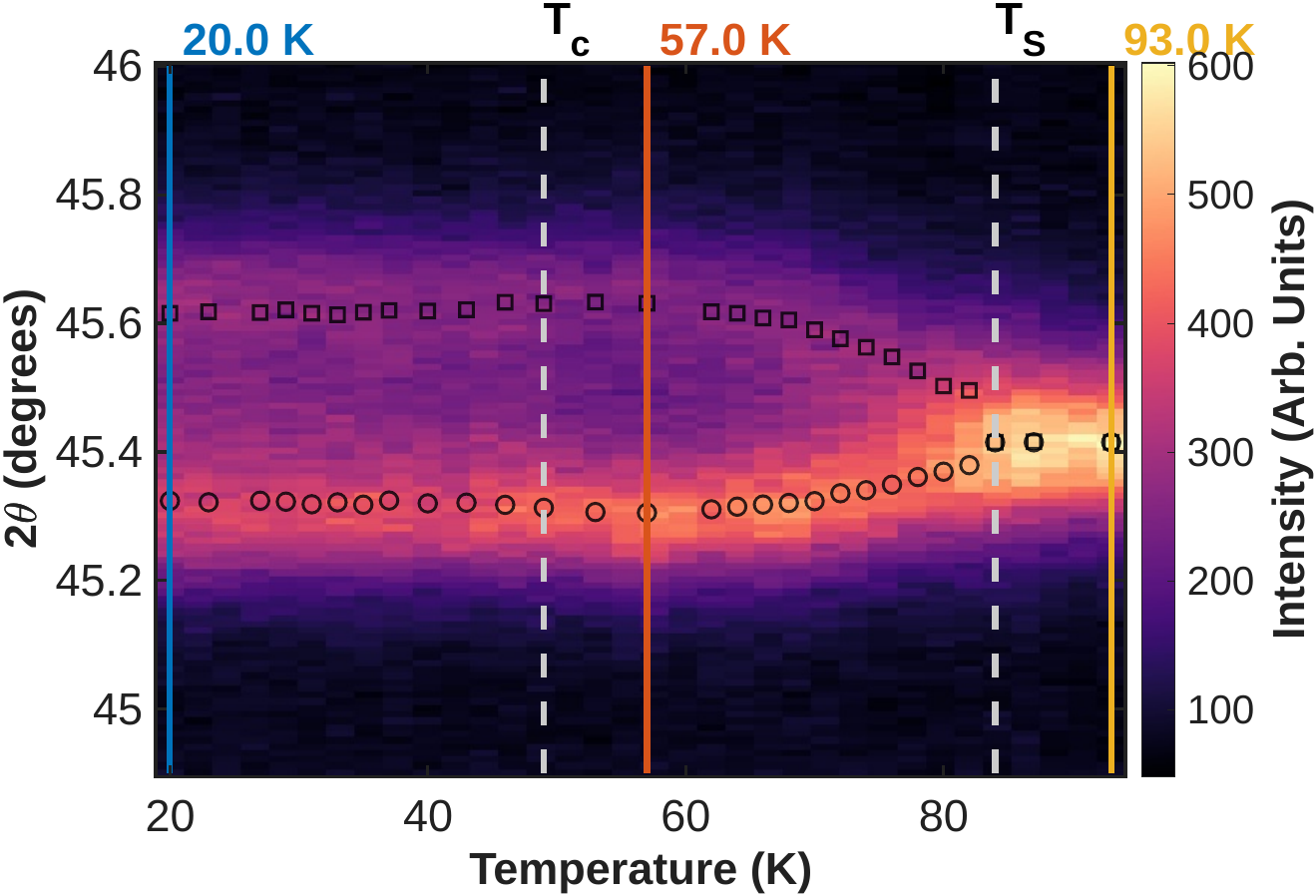} \\
			\textbf{(a) Smartlab} \\
			\includegraphics[width=\linewidth]{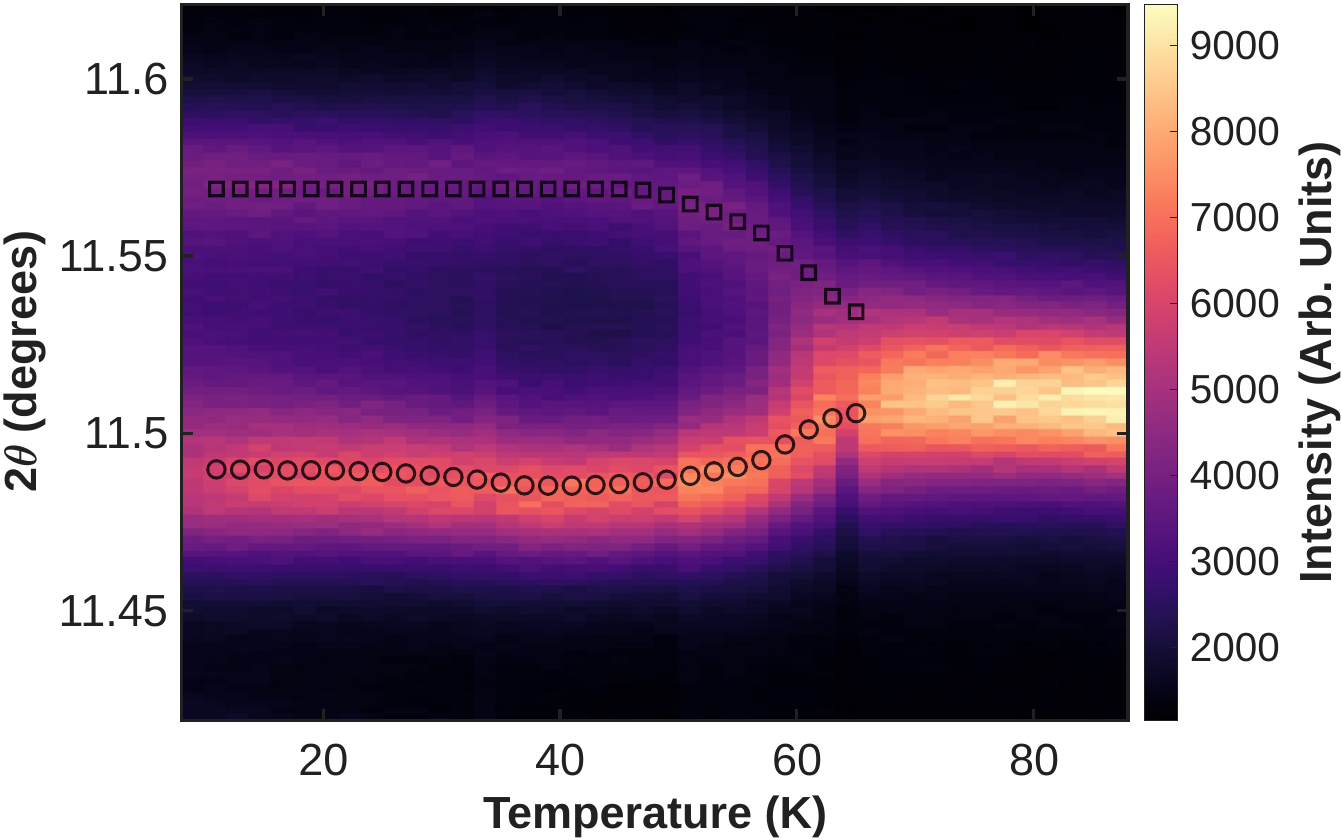} \\
			\textbf{(b) ESRF}
		}
		&
		\parbox{0.42\textwidth}{
			\centering
			\includegraphics[width=0.9\linewidth]{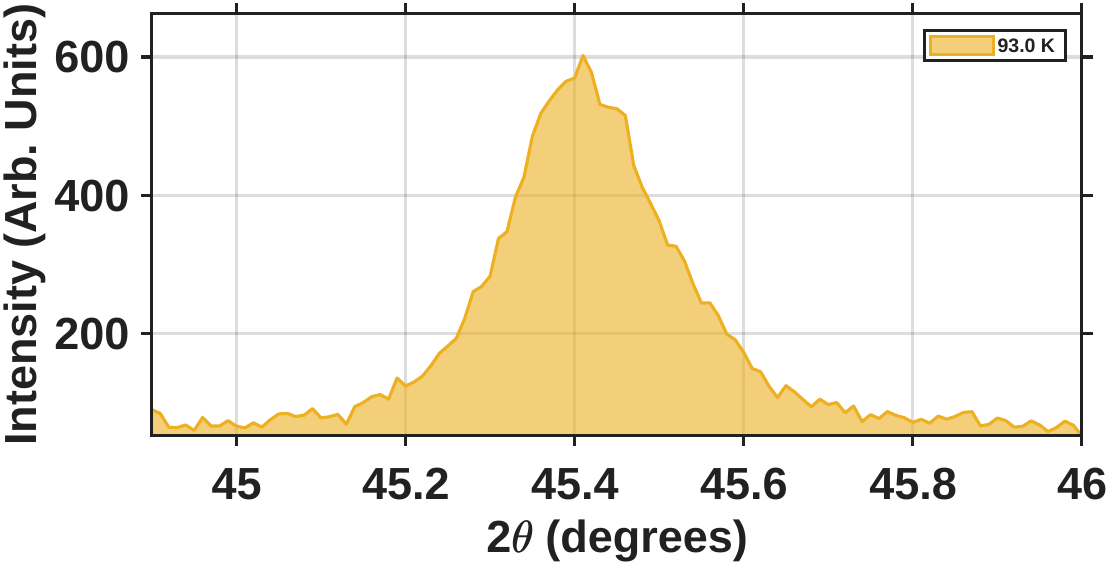} \\
			\textbf{(c) 93 K (R$\overline{3}$)} \\
			\includegraphics[width=0.9\linewidth]{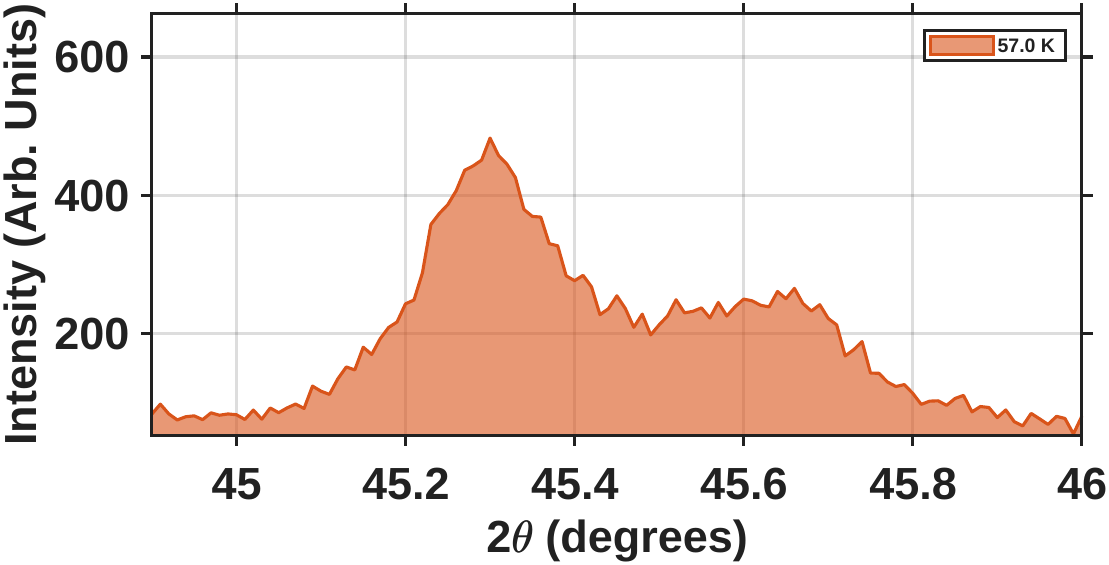} \\
			\textbf{(d) 57 K (P$\overline{1}$)} \\
			\includegraphics[width=0.9\linewidth]{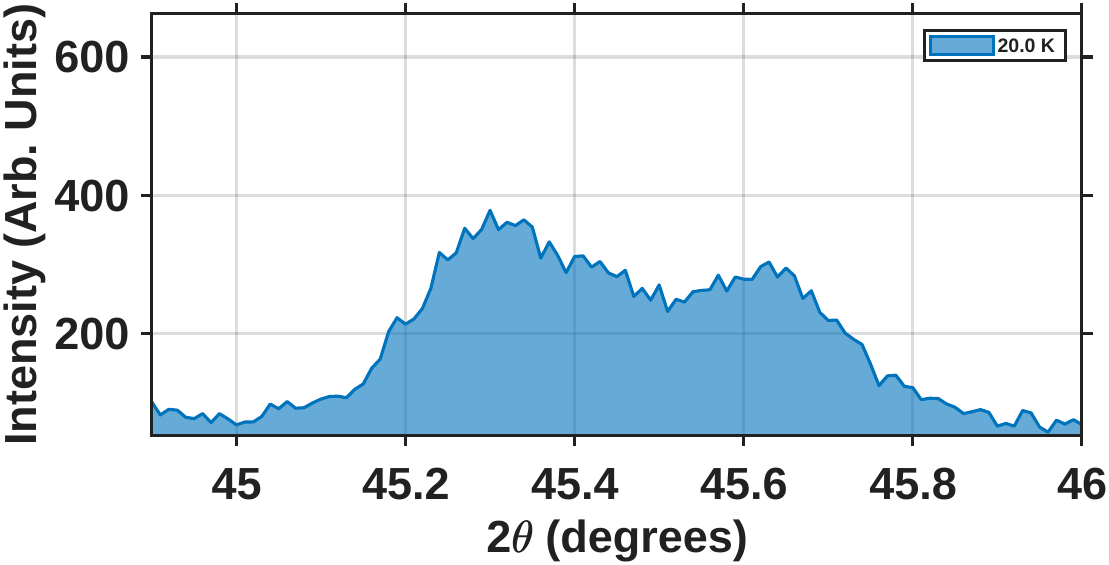} \\
			\textbf{(e) 20 K (P$\overline{1}$)}
		}
	\end{tabular}
	\caption{Structural transition in crushed VI$_3$ single crystals. $(a-b)$ Temperature evolution of the (3,0,0) Bragg peak showing the $R\overline{3}$ to $P\overline{1}$ transition at $\sim$ 80 K (Smartlab) and $\sim$ 70 K (ESRF). $(c-e)$ Representative diffraction patterns showing: $(c)$ high-temperature R$\overline{3}$ phase, $(d)$ intermediate $P\overline{1}$ phase, and $(e)$ fully split low-temperature $P\overline{1}$ pattern. The $\sim$ 50 K anomaly corresponds to magnetostriction at T$_C$.}
	\label{fig:INTENSITY}
\end{figure*}

We begin by examining the structural and magnetic phase transitions in VI$_3$ as observed in our single-crystal samples applying neutron diffraction using triple-axis spectrometers.  Fig. \ref{fig:neutron_elastic} illustrates the temperature dependence of the $\vec{Q}=(1,1,0)$ Bragg peak, with peak intensity as a function of temperature shown in Fig. \ref{fig:neutron_elastic} $(a)$ (using MACS) and A3 scans (using SPINS) through the Bragg peak displayed in Fig. \ref{fig:neutron_elastic} $(b)$.  The temperature dependent intensity shown in Fig. \ref{fig:neutron_elastic} $(a)$ displays anomalies at $\sim$ 80 K and again at $\sim$ 50 K.  These two temperatures are consistent with the reported high temperature structural T$_{S}$ and the lower temperature ferromagnetic T$_{c}$ transition.~\cite{Kong19:31,Gati19:100,Tian19,Son19:99,Liu20:1} 

\begin{figure*}
	\centering
	\begin{tabular}{cccc}
		\parbox{0.34\textwidth}{
			\centering
			\includegraphics[width=\linewidth]{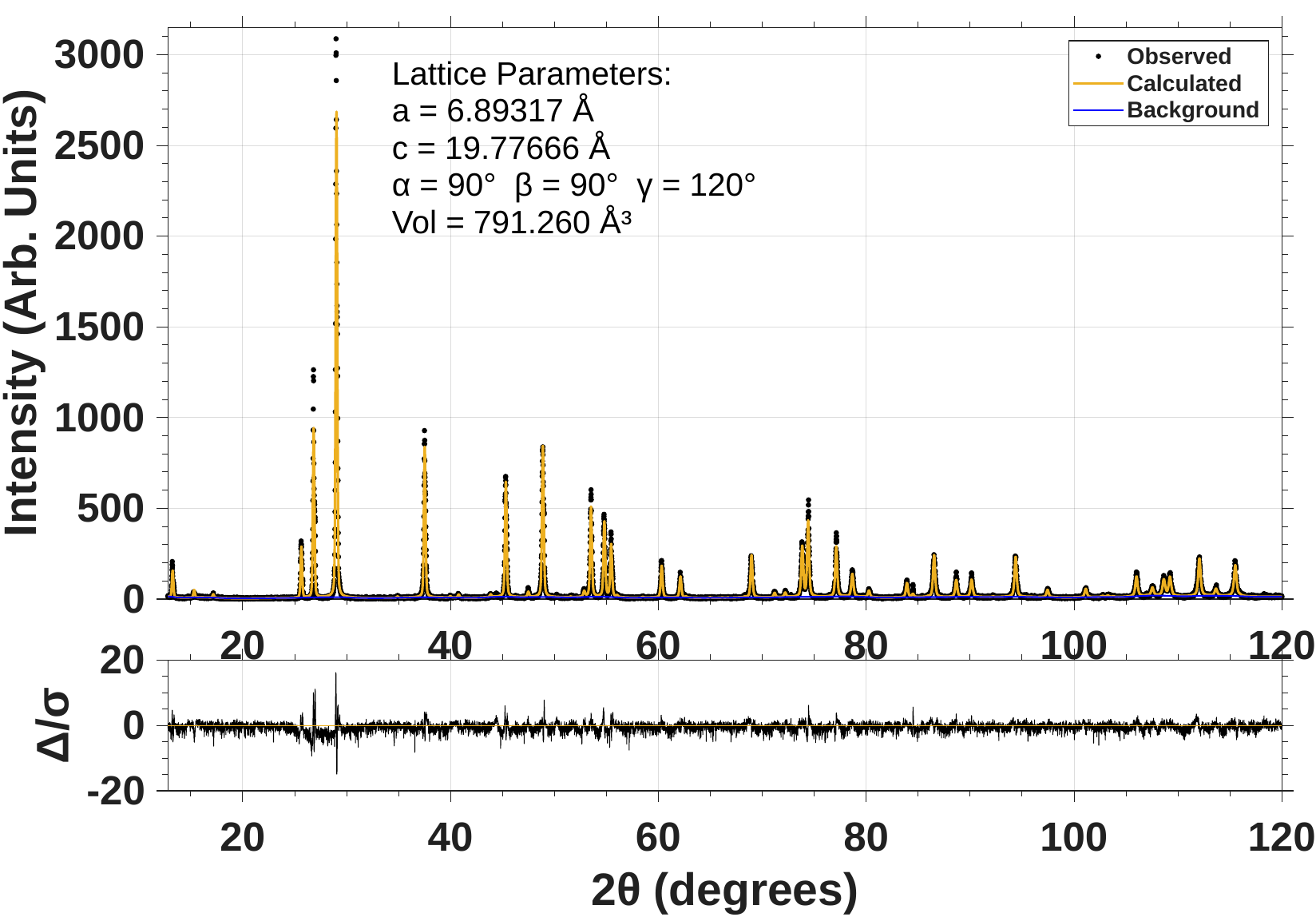} \\
			\textbf{(a) Le Bail Refinement 100 K (R$\overline{3}$)} \\
			\vspace{0.2cm}
			\includegraphics[width=\linewidth]{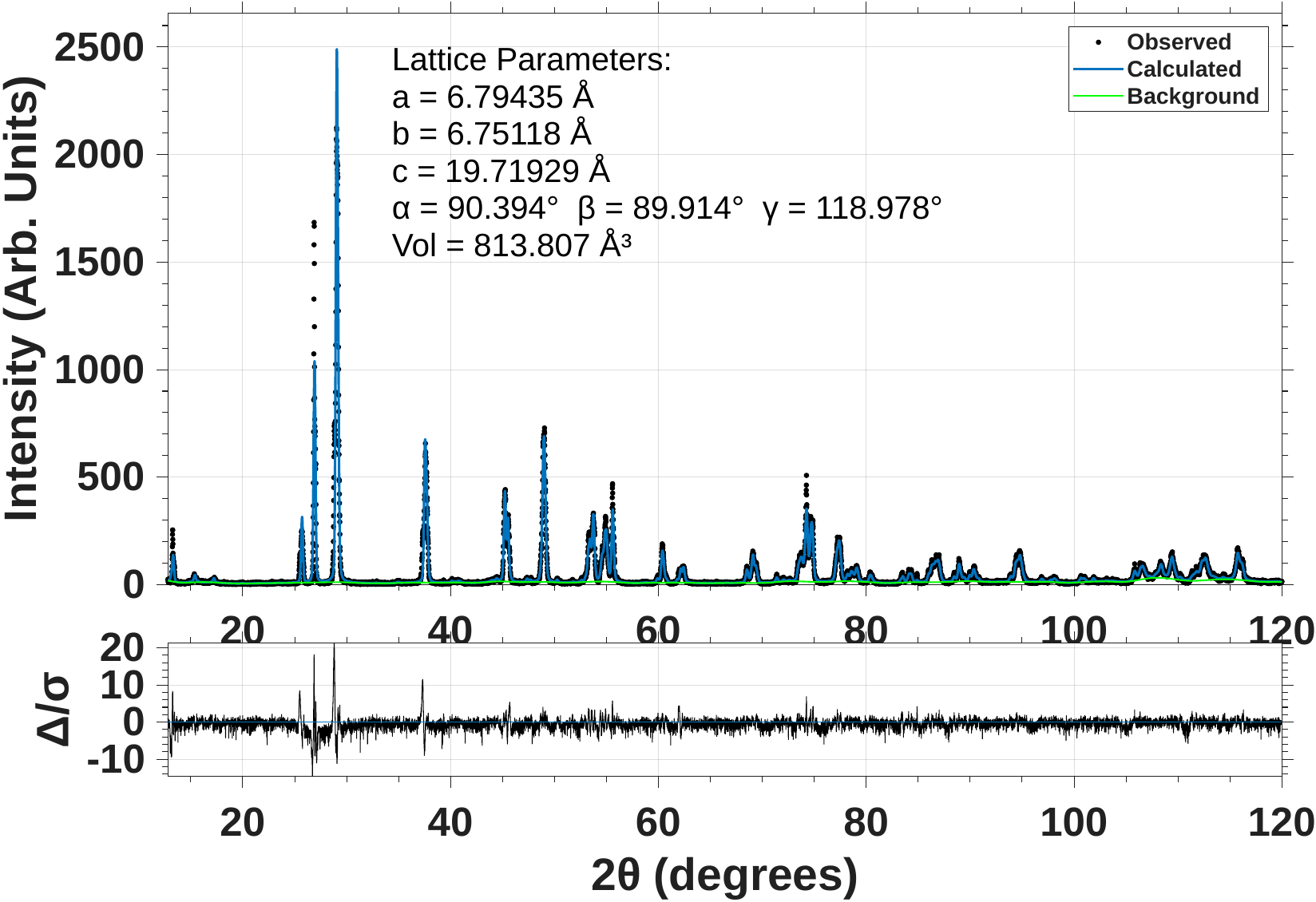} \\
			\textbf{(b) Le Bail Refinement 15 K (P$\overline{1}$)}
		}
		&
		\parbox{0.21\textwidth}{
			\centering
			\includegraphics[width=\linewidth]{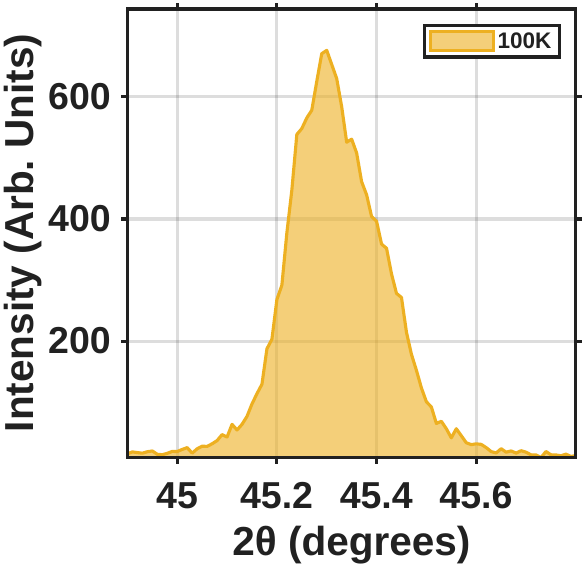} \\
			\textbf{(c) Sample A (100 K)} \\
			\vspace{0.8cm}
			\includegraphics[width=\linewidth]{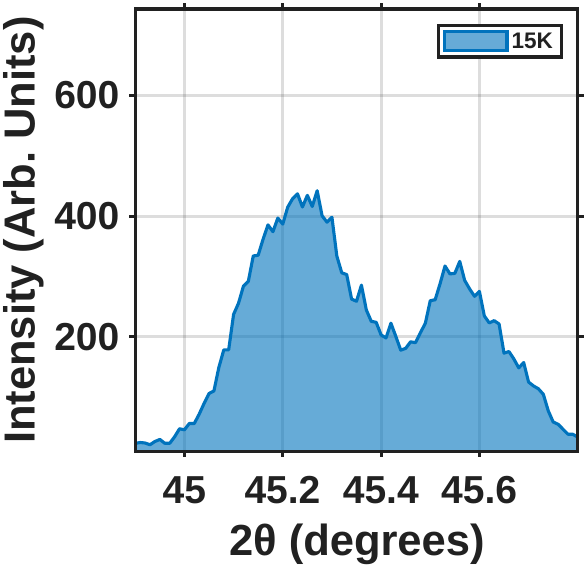} \\
			\textbf{(d) Sample A (15 K)}
		}
		&
		\parbox{0.21\textwidth}{
			\centering
			\includegraphics[width=\linewidth]{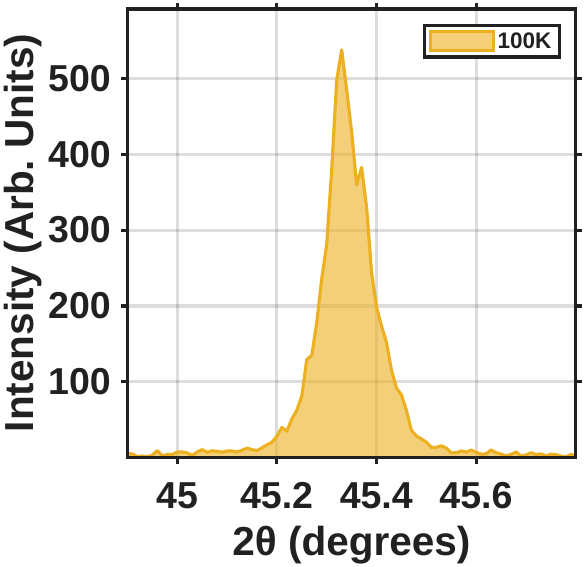}\\
			\textbf{(e) Sample B (100 K)} \\
			\vspace{0.8cm}
			\includegraphics[width=\linewidth]{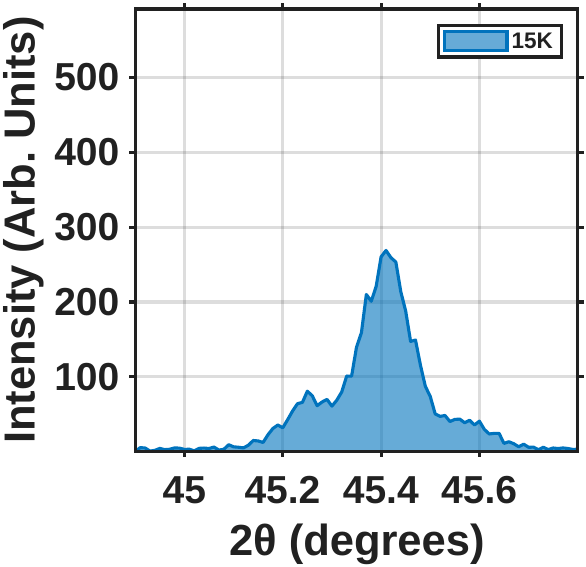} \\
			\textbf{(f) Sample B (15 K)}
		}
		&
		\parbox{0.21\textwidth}{
			\centering
			\includegraphics[width=\linewidth]{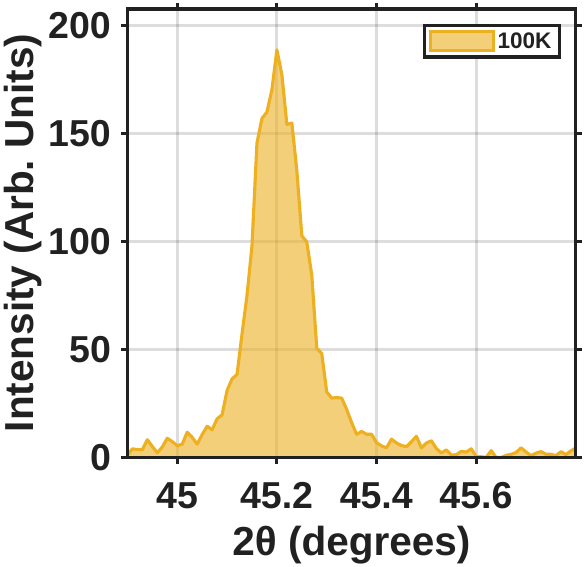} \\
			\textbf{(g) Sample C (100 K)} \\
			\vspace{0.8cm}
			\includegraphics[width=\linewidth]{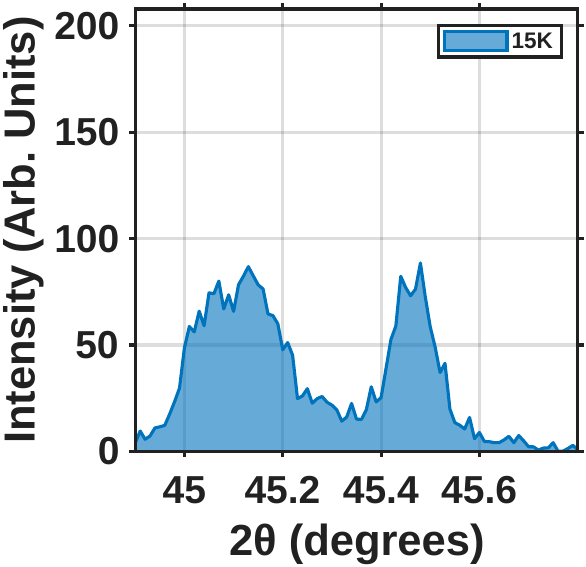} \\
			\textbf{(h) Sample C (15 K)}
		}
	\end{tabular}
	\caption{
		Structural characterization of VI$_{3}$ powder samples. $(a-b)$ Le Bail refinements confirming $R\overline{3}$ to $P\overline{1}$ transition. $(c-h)$ Sample-dependent behavior: 
		Sample A (400°C quench): $\sim$ 80 K transition to $P\overline{1}$.
		Sample B (500°C slow-cool): $\sim$ 30 K transition with weak splitting.
		Sample C (400°C slow-cool): Coexistence of both transitions ($\sim$ 80 K main + $\sim$ 30 K secondary splitting).
		The variable transition temperatures suggest stacking-dependent polymorphism.
	}
	\label{fig:samples}
\end{figure*}

To further investigate the temperature dependent properties of the structure in our single crystal samples, we apply powder diffraction on crushed single crystals in Fig. \ref{fig:INTENSITY}.  Fig. \ref{fig:INTENSITY} $(a)$ illustrates a false color map of the splitting of the $\vec{Q}=(3,0,0)$ Bragg peak indexed in the high temperature $R\overline{3}$ space group.  The data was taken in reflection Bragg-Bretano geometry using a PheniX displex on a Rigaku Smartlab ($\lambda$=1.54 \AA). The reflection geometry on lightly ground single crystal data illustrates a splitting of the Bragg peak at T$_{S}\sim$ 80 K with illustrative $2\theta$ scans displayed in Fig. \ref{fig:INTENSITY} $(c-e)$.  We discuss the low-temperature space group below based on Le Bail fits to the unit cell shape.  A further experiment was performed on ID22 (ESRF) on a capillary filled with crushed single crystals loaded in an argon filled glovebox and $2\theta$ scans through the same $\vec{Q}=(3,0,0)$ Bragg peak are shown in Fig. \ref{fig:INTENSITY} $(b)$.  Similar to the Smartlab measurement, this measurement also shows a structural transition, but a reduced temperature of $\sim$ 70 K.  Given concerns due to thermal conductivity of the glass capillary and synchrotron beam heating, we consider the correct structural transition to be given by the Smartlab data with the PheniX displex in reflection geometry.  We tested for the possibility of mismatched temperatures on the PheniX displex and actual sample temperature through measurements of CoCl$_{2}$ finding a transition temperature consistent with the reported $\sim$ 25 K transition.~\cite{Moses80:13}  We therefore conclude that the Smartlab PheniX system provides the more accurate transition temperature with T$_{S}\sim$ 80 K.

Both Figs. \ref{fig:INTENSITY} $(a-b)$ display a single structural transition which we assign to T$_{S}\sim$ 80 K with no observable structural transition seen near $\sim$ 30 K.  Both the Smartlab and ESRF data show a subtle change in the low temperature lattice constants near $\sim$ 50 K, coinciding with the ferromagnetic transition found on the same single crystal samples in Fig. \ref{fig:neutron_elastic} $(a)$ measured with neutrons.  The change in $2\theta$ position below ferromagnetic T$_{C}$ corresponds to a lattice constant change of order ${{\Delta d} \over {d}} \sim 10^{-3}$.  This is indicative of sizeable magnetostriction at the ferromagnetic transition especially when compared to conventional ferromagnets such as Ni~\cite{Lee71:326} and Fe~\cite{Lacheisserie83:31} which have magnetostriction coefficients of $\lambda \sim$ 10$^{-5}$.  Such a structural change at T$_{c}$ was previously also reported in the context of refinements to a monoclinic unit cell reporting a sizeable change in the value of $\beta$ at the magnetic Curie temperature.~\cite{Kratochvilova22:34}

We discuss this further below in the context of spin-orbit coupling and the magnetic excitations.  We therefore conclude that our single crystals display a single structural transition at T$_{S} \sim$ 80 K and a ferromagnetic transition at T$_{C} \sim$ 50 K.  While our diffraction data, based on Le Bail fits to the unit cell and supported by group theoretical analysis, do not yield information on the atomic positions of the two sites, we show below that spectroscopic measurements provide strong evidence for two distinct V$^{3+}$ sites in VI$_3$.

\subsection{Powder samples}

We now discuss x-ray diffraction results for our 3 different powder samples synthesized as discussed above in the experimental section above. Given issues regarding preferred orientation, we present the results based on Le Bail fits~\cite{Bail05:20} performed using \texttt{GSAS-II}~\cite{Toby13:46} to the powder data where the peak positions ($2\theta$) are fit to refine a unit cell shape.  The comparative results for the three different powder samples is displayed in Fig. \ref{fig:samples}.  We first present diffraction data and then discuss these in the context of the literature, lastly speculating as to the origin of the differences.

Our structural analysis reveals three distinct behavioral responses in the three synthesized VI$_3$ powder samples. Sample A (400$^\circ$C quenched) exhibits identical behavior to the results presented above for single crystals, displaying only a $\sim$ 80 K structural transition evidenced by splitting of the $\vec{Q}$=(3,0,0) Bragg peak.  Le Bail fits to the powder data shown in Fig. \ref{fig:samples} $(a-b)$ suggest a transition from a high temperature $R\overline{3}$ space group to a low temperature space group with lower symmetry.  We have fit this to a triclinic unit cell of symmetry $P\overline{1}$.  While it is always possible to fit a unit cell to a triclinic space group, we discuss the motivation for this below in the context of group theory.   The splitting is illustrated in  Fig. \ref{fig:samples} $(c,d)$ and akin to the data above, no measurable difference in the data was observed between 15 K and 50 K in the peak splitting indicative that any further structural transition is not observable.

\begin{figure}[!htbp]
	\centering
	\includegraphics[width=0.9\linewidth]{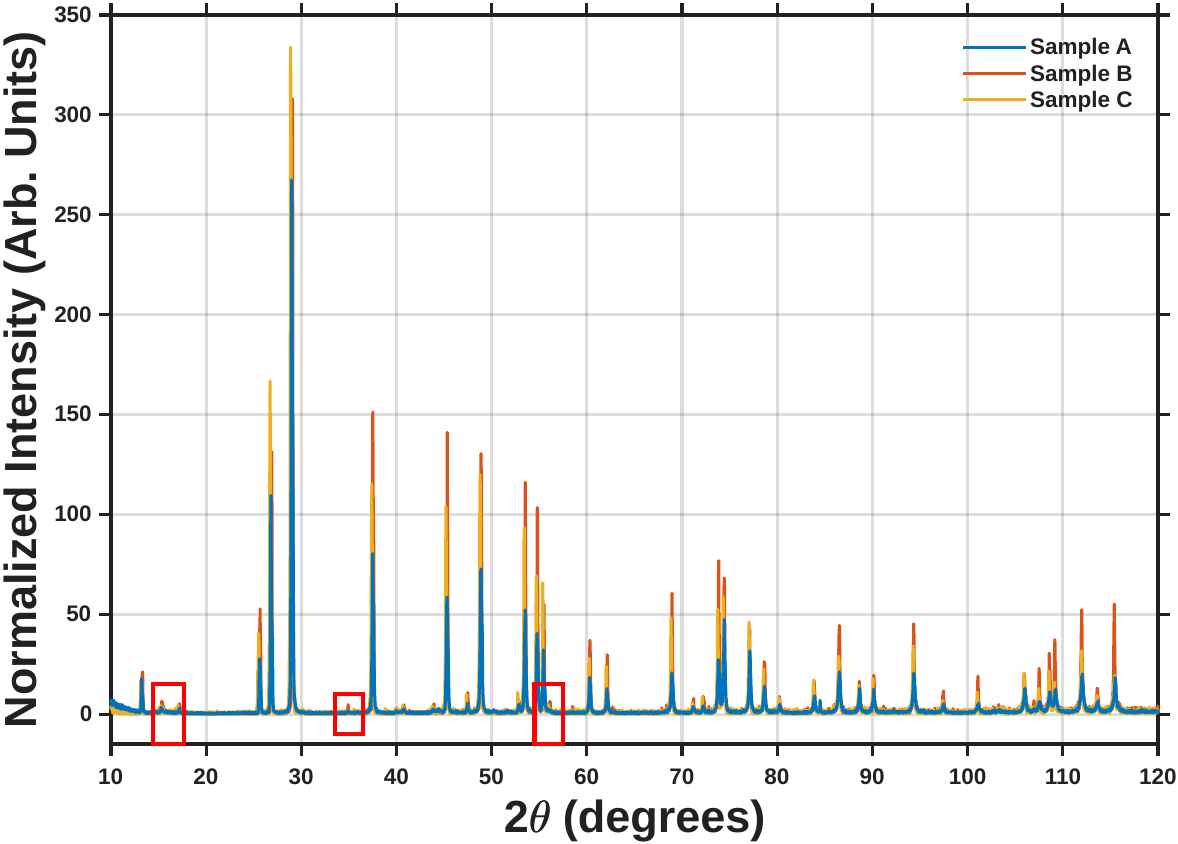} \\
	\textbf{(a) Comparison of all samples at 100 K} \\[1ex]
	\includegraphics[width=0.9\linewidth]{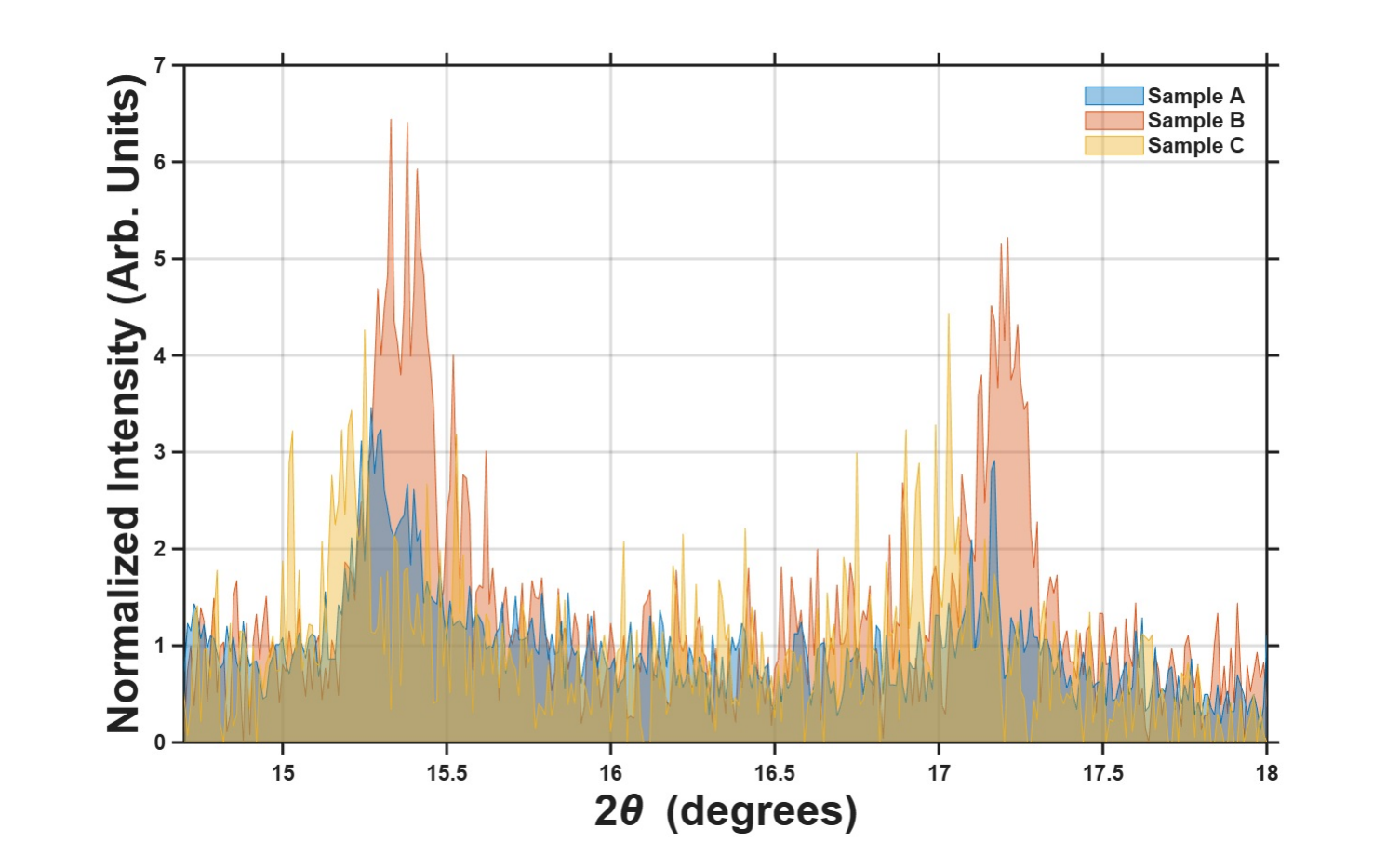} \\
	\textbf{(b) Sample B peaks at (1 0 1) and (1 0 -2)} \\[1ex]
	\includegraphics[width=0.9\linewidth]{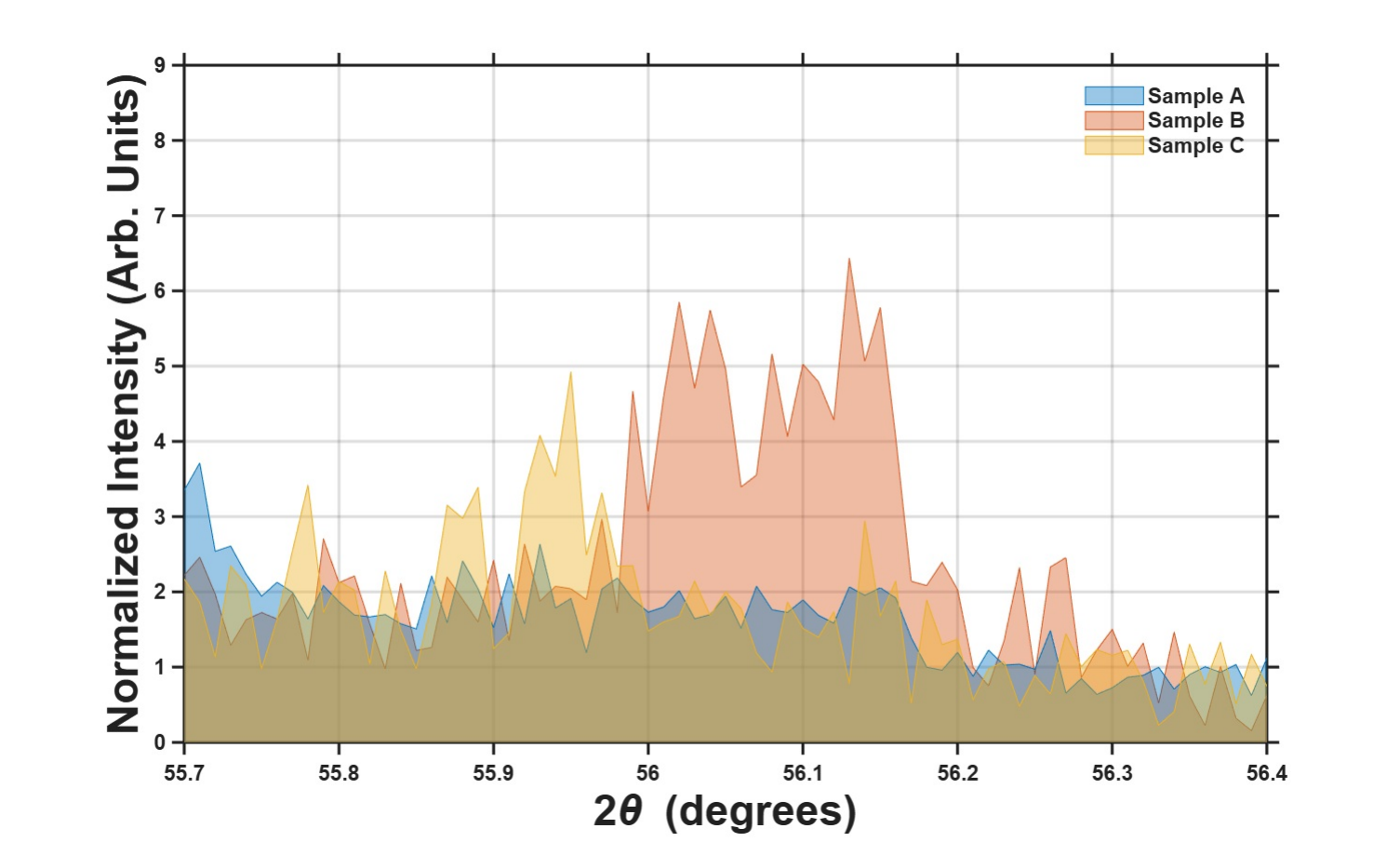} \\
	\textbf{(c) Sample B peak at (0 0 12)}
	\caption{
		Diffraction patterns revealing sample-specific structural features. 
		$(a)$ Comparative 100 K patterns showing overall phase consistency, with red boxes highlighting distinctive features in Sample B. 
		$(b)$ Two characteristic peaks at (1 0 1) and (1 0 -2) unique to Sample B, suggesting possible stacking faults. 
		$(c)$ The (0 0 12) reflection appearing exclusively in Sample B. 
	}
	\label{fig:peak}
\end{figure}

Sample B (500$^\circ$C slow-cooled) demonstrates fundamentally different characteristics (Fig.~\ref{fig:samples} $e-f$ and Fig.~\ref{fig:peak}). This sample shows only partial peak splitting at $\sim$ 30 K, accompanied by high temperature additional reflections over other samples that we index based on the $R\overline{3}$ high temperature space group to (1 0 1), (1 0 -2), and (0 0 12).  Notably, this sample does not display a $\sim$ 80 K structural transition observed in the samples discussed above and this partial splitting was found to be onset at a lower temperature of $\sim$ 30 K. 

\begin{figure*}[!htbp]
	\centering
	\begin{tabular}{cc}
		\parbox{0.47\textwidth}{
			\centering
			\includegraphics[width=\linewidth]{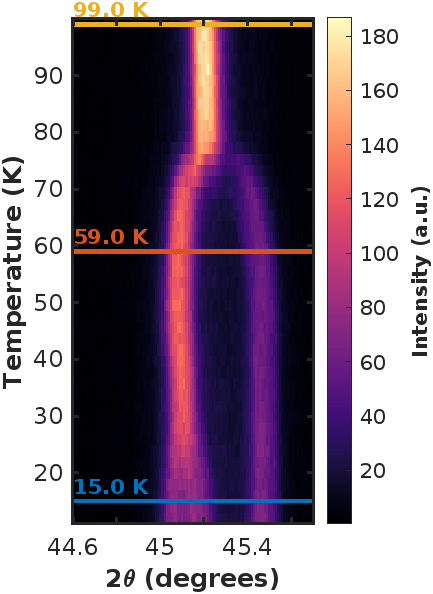} \\
			\textbf{(a) Temperature dependence (Sample C)}
		}
		&
		\parbox{0.42\textwidth}{
			\centering
			\includegraphics[width=0.9\linewidth]{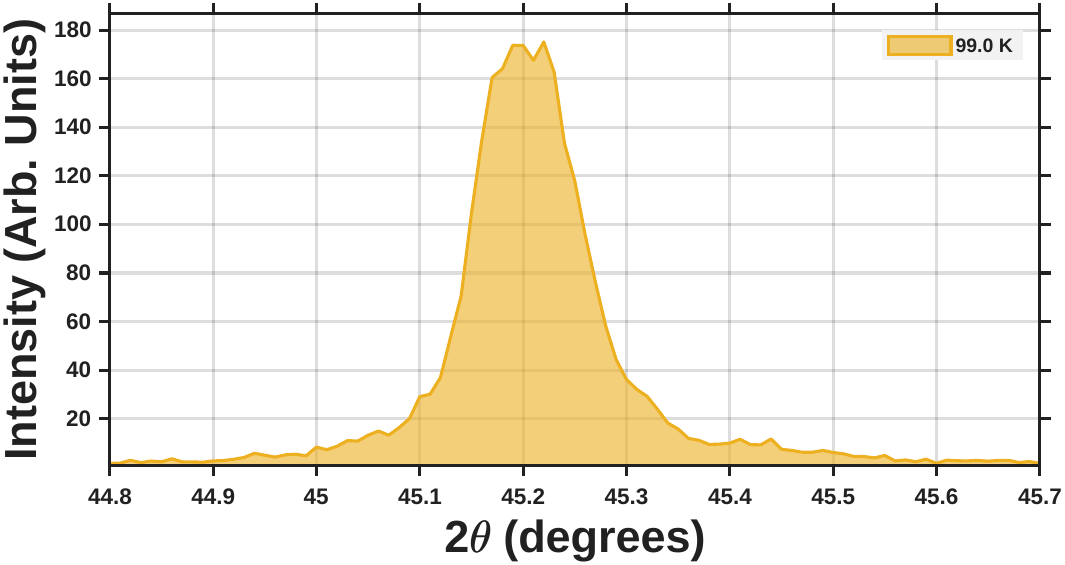} \\
			\textbf{(b) R$\overline{3}$ phase (99 K)} \\
			\includegraphics[width=0.9\linewidth]{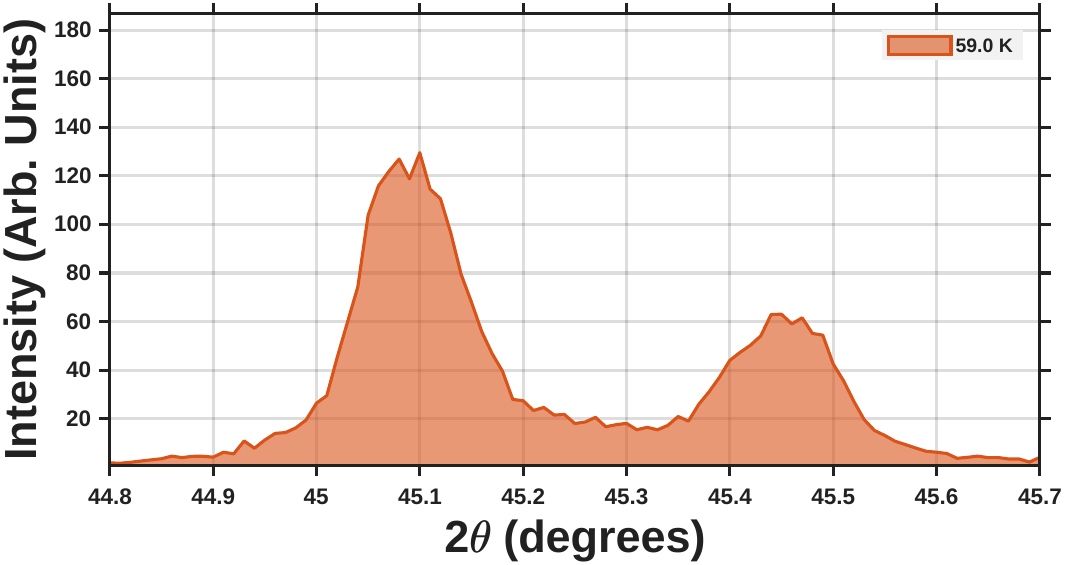} \\
			\textbf{(c) P$\overline{1}$ phase (59 K)} \\
			\includegraphics[width=0.9\linewidth]{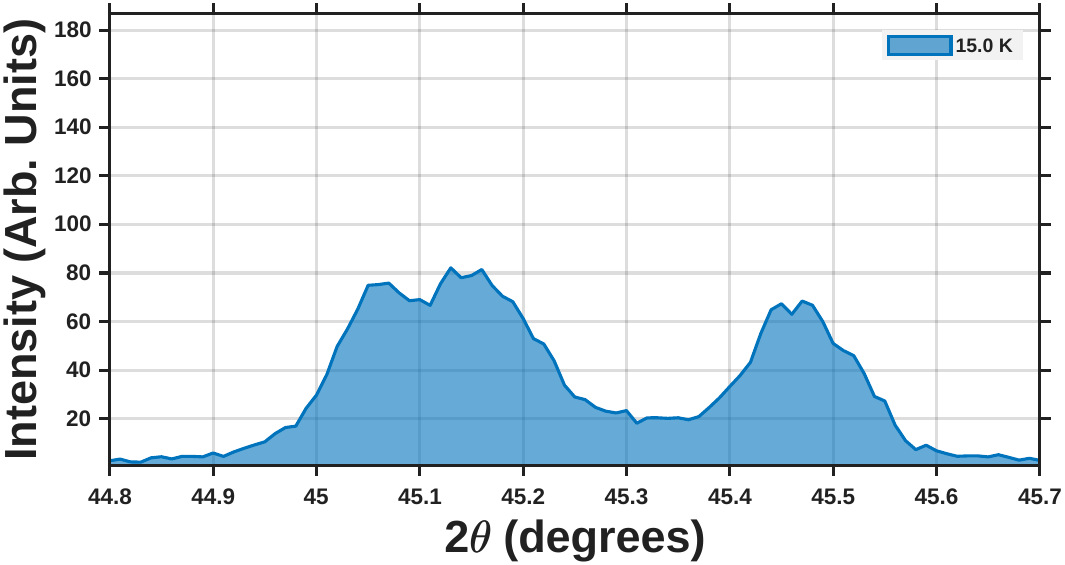} \\
			\textbf{(d) P$\overline{1}$+stacking (15 K)}
		}
	\end{tabular}
	\caption{
		Temperature-dependent structural evolution in Sample C. $(a)$ Intensity map showing two distinct splitting events at $\sim$80 K and $\sim$30 K. $(b)$ High-temperature phase with single Lorentzian peaks. $(c)$ Intermediate phase showing primary splitting. $(d)$ Low-temperature phase with additional peak splitting. The complex evolution suggests competing structural motifs emerging from the specific 400$^\circ$C slow-cool synthesis conditions.
	}
	\label{fig:INTENSITY2}
\end{figure*}

Sample C (400$^\circ$C slow-cooled) presents hybrid behavior (Fig.~\ref{fig:samples} $g-h$ and Fig.~\ref{fig:INTENSITY2}), exhibiting both the characteristic $\sim$ 80 K transition and additional $\sim$ 30 K peak broadening suggestive of a further peak splitting indicative of a second lower temperature structural transition. In Fig.~\ref{fig:INTENSITY2} $(a)$, we observe clear peak splitting developing at two distinct temperatures. The subsequent panels reveal the detailed evolution: Fig.~\ref{fig:INTENSITY2} $(b)$ shows the high-temperature $R\overline{3}$ phase with single peaks, Fig.~\ref{fig:INTENSITY2} $(c)$ demonstrates the primary splitting into two peaks with a scan at 59 K of the $\vec{Q}$=(3,0,0) Bragg peak, and Fig.~\ref{fig:INTENSITY2}$(d)$ reveals the secondary splitting where the first lower $2\theta$ peak from Fig.~\ref{fig:INTENSITY2} $(c)$ further divides into two components below 30 K. This contrasts sharply with the behavior seen in Fig.~\ref{fig:INTENSITY} for crushed single crystals, where only a single observable splitting occurs at $\sim$80 K. The distinct behavior of Sample C likely originates from the specific synthesis conditions (400$^\circ$C with slow cooling), which may promote partial transformation of VI$_3$ sites while preserving the overall framework.

Contrary to earlier reports~\cite{Tian19,Dolezal19:3,Kratochvilova22:34} suggesting monoclinic intermediate phases or multiple transitions near $\sim$30 K, our high-resolution x-ray data reveal no evidence of an additional monoclinic $C2/m$ symmetry at any temperature. We therefore suggest that the structural distortion in pure VI$_3$ occurs  exclusively at $\sim$80 K without subsequent symmetry changes.  We now discuss this point further in the context of group theory below.

\subsection{Symmetry and Group Theory}

\begin{table}[h!]
	\centering
	\begin{tabular}{@{}l@{\hspace{20pt}}l@{}}
		\hline
		\textbf{Space Group} & \textbf{Symmetry Operations} \\

		\( R\overline{3} \) & 
		\begin{tabular}[t]{@{}l@{}}
			\( (x, y, z) \) \\
			\( (-y, x - y, z) \) \\
			\( (-x + y, -x, z) \) \\
			\( (-x, -y, -z) \) \\
			\( (y, -x + y, -z) \) \\
			\( (x - y, x, -z) \) \\
		\end{tabular} \\

		\( P\overline{1} \) & 
		\begin{tabular}[t]{@{}l@{}}
			\( (x, y, z) \) \\
			\( (-x, -y, -z) \) \\
		\end{tabular} \\

		\( P1 \) & 
		\begin{tabular}[t]{@{}l@{}}
			\( (x, y, z) \) \\
		\end{tabular} \\
		\hline
	\end{tabular}
	\caption{Symmetry Operations in Fractional Coordinates ($x, y, z$)}
	\label{Table:symmetry_elements}
\end{table}

Using group-theoretical symmetry analysis, we systematically examined (using the Bilbao crystallographic server~\cite{Aroyo06:221,Aroyo06:A62} programs \texttt{MAXSUB} and \texttt{SUBGROUPGRAPH}~\cite{Ivantchev00:33,Ivantchev02:35}) the possible phase transitions from the $R\overline{3}$ structure based on group-subgroup relations~\cite{Stokes84:30}.  With a structural symmetry breaking, it is conventionally expected that the transition will remove a symmetry element and therefore the lower temperature will have fewer symmetry elements in its group, forming a subgroup of the parent high temperature space group.   The maximal subgroups of the $R\overline{3}$ space group include the triclinic $P\overline{1}$ and space groups with threefold symmetry indexed to either a primitive or rhombohedral unit cell.  There is no monoclinic subgroup or supergroup.  The symmetry elements of the $R\overline{3}$, $P\overline{1}$, and the $P1$ space groups are tabulated in Table \ref{Table:symmetry_elements}.

Our analysis finds a splitting of the Bragg peaks and a breaking of the threefold symmetry indicated by a splitting of the $\vec{Q}$=(3,0,0) Bragg peak discussed above resulting in a complex pattern that fits to a transition to $P\overline{1}$ (or $P1$).  This indicates clear evidence of a symmetry lowering to the triclinic $P\overline{1}$ phase around $\sim$ 80~K. We note that a similar transition has been reported in VBr$_{3}$.~\cite{Gu24:110}  In contrast, transitions to a rhombohedral phase can be ruled out due to the complexity of the powder diffraction pattern. Contrary to earlier reports~\cite{Tian19, Dolezal19:3, Kratochvilova22:34} suggesting a monoclinic intermediate phase or a second transition near $\sim$ 30 K, we find no such features and such transitions are furthermore not supported by group theory.  Furthermore, with a transition to a triclinic unit cell, there are no further sub groups (other than the transition for $P\overline{1} \rightarrow P1$ which only removes the inversion element $(-x,-y,-z)$) and therefore we would not expect to see peak splitting in the powder pattern as displayed in Fig. \ref{fig:INTENSITY2}.

As indicated above, a transition from $R\overline{3}$ to $P\overline{1}$ would require the breaking of all rotational symmetry elements of the $R\overline{3}$ space group.  This would introduce the possibility of this transition being first-order.  We note that while our data is largely consistent with a continuous transition (second-order), there is evidence near the structural transition in our data and also those reported in Ref. \onlinecite{Dolezal19:3} of coexistence which would support it being first-order.  The existence of a first-order transition would support our group theoretical analysis.

The $\sim$ 30 K structural transition features in Samples B and C likely originate from competing structural motifs rather than true phase transitions. We speculate on the origin and attribute to several possible mechanisms. First, we propose that Moir\'e stacking of the layers in VI$_{3}$ maybe the origin of the $\sim$ 30 K transition.  This is based on the observation of additional $(H,K,L)$ peaks in the diffraction pattern (Fig. \ref{fig:peak}) which index off large values of $L$ indicative of more ordered stacking than present in samples studied which lacked this transition. This may result in a stacking sequence that reflects more the $AAA$ stacking sequences found in VI$_2$~\cite{Kuindersma79:30} which has been reported to have a magnetic transition at $\sim$20-30 K applying neutron diffraction.  VI$_3$ in this regard differs because of the the $ABC$ stacking arrangement.  It is noteworthy that related compounds (Cr,Br)I$_{2}$~\cite{Schneeloch24:109,Schneeloch25:111} undergo magnetic transitions in a similar temperature range. Multiple structural polymorphs, with only subtle static structural features to distinguish, have been suggested in these materials based on differing stacking sequences.~\cite{Zhang22:105}  

A second possibility for the $\sim$ 30 K transition is temperature-dependent competition between V-V and I-I interactions that could drive local site transformations from VI$_3$-like to VI$_2$-like coordination. We note that VI$_{2}$ and its related transition metal ion counterparts differ from VI$_{3}$ not only in terms of the stacking of the Van der Waals layers, but in the approximately hexagonal in-plane structure rather than the honeycomb in VI$_{3}$.  We discuss this below in the context of the magnetic dynamics.   Finally and a possible third, strain fields from coexisting stacking types may induce peak broadening without symmetry change.  Such a situation occurs in disordered relaxor ferroelectrics such as PZN-PT~\cite{Stock04:69,Koo02:65,Stock07:76} where the internal strain results in Bragg peak broadening rather than a resolvable splitting of the Bragg peaks.

These interpretations suggest that no true symmetry change occurs below $\sim$80 K - the additional splittings represent structural modulations within the established P$\overline{1}$ framework. We speculate that the resulting Moir\'e patterns from competing stackings~\cite{Yao24:15} may enable novel low temperature states in this van der Waals magnet, while the robust $\sim$80 K transition provides a unified framework for understanding VI$_3$'s fundamental physics.  In this context it is interesting to note that temperature dependent specific heat measurements~\cite{Dolezal19:3} do not display strong anomalies at $\sim$ 30 K in comparison to those measured at the $\sim$ 50 K and $\sim$ 80 K ferromagnetic and structural transitions.

\section{Magnetic dynamics}

We now discuss the connection between structural and magnetic properties and first argue that spin-orbit coupling is an important and relevant energy scale in VI$_{3}$ and use this to model the low-energy magnetic dynamics probed with neutron scattering. Above, applying x-ray diffraction we investigated the temperature dependent structural properties of VI$_{3}$ and advocated for the presence of a single structural transition at T$_{S}\sim$ 80 K followed by a lower temperature ferromagnetic transition at $\sim$ 50 K. X-ray diffraction also finds a measurable change in the lattice constant (reflected by the $2\theta$ position of the $\vec{Q}$=(3,0,0) Bragg peak) at the ferromagnetic transition that is considerably larger than typical ferromagnets such as Ni or Fe.  This is strongly suggestive of a coupling between the lattice and the magnetic spins.  A direct pathway for coupling the lattice and spin degrees of freedom is provided by the spin-orbit interaction, represented as $\sim \vec{l} \cdot \vec{S}$. The relevance of spin-orbit coupling is further underscored by a high-temperature structural transition, which is anticipated based on the Jahn Teller theorem in the presence of orbital degeneracy such as that expected for V$^{3+}$ ions in an ideal octahedral crystal field environment~\cite{Gehring75:38}.  

We suggest that the presence of a spontaneous symmetry breaking structural transition, discussed above in the introduction, is a result of an orbital degree of freedom in VI$_{3}$ and consequently the spin-orbit interaction being a relevant term in the magnetic Hamiltonian.  V$^{3+}$ ions in VI$_{3}$ reside in an octahedral crystalline electric field which breaks the fivefold $3d$ orbital degeneracy.  In terms of the strong crystal field basis, such a Coulomb field results in lower energy triply degenerate $|t_{g}\rangle$ states and a higher energy doubly degenerate $|e_{g}\rangle$ levels.  Populating the low-energy $|t_{g}\rangle$ levels with 2-$d$ electrons results in an orbital degeneracy that exactly maps onto an $l_{eff}=1$.~\cite{Cowley13:88,Sarte18:98,Sarte19:100,Sarte20:102,Popescu25:134,McClure:book,Abragam:book}

Beyond localized crystal field single-ion arguments for the presence of an orbital degree of freedom and our diffraction results discussed above, there are a number of other results that support the importance of orbital filling in the structural and electronic properties of VI$_{3}$. X-ray dichorism illustrates the presence of an orbital contribution to the statically ordered magnetic moment~\cite{Hovancik23:23} and x-ray spectroscopy supports the presence of an unquenched orbital degree of freedom.~\cite{Vita22:22,Vita24:24} Strain tuneability of the magnetic anisotropy~\cite{Zhang22:22} further supports a link between magnetism and structure that spin-orbit coupling facilitates~\cite{Yosida:book}.  Density functional calculations, which accurately reproduce the measured magnetic moment, find orbital contributions to the magnetism~\cite{Sandratskii21:103,Hao21:38} and an unquenched orbital moment~\cite{Xu25:36}. This has also been suggested based on Hall measurements~\cite{Zhang21:127} and terahertz~\cite{Hovancik22:13} measurements.  In our analysis below of the magnetic excitations we evaluate the role of an orbital degree of freedom has on the magnetic excitations and also its link with the structure and our findings of a triclinic distortion above.

\begin{figure}
	\centering
	\includegraphics[width=95mm,trim=2.0cm 5.25cm 1.5cm 5.0cm,clip=true]{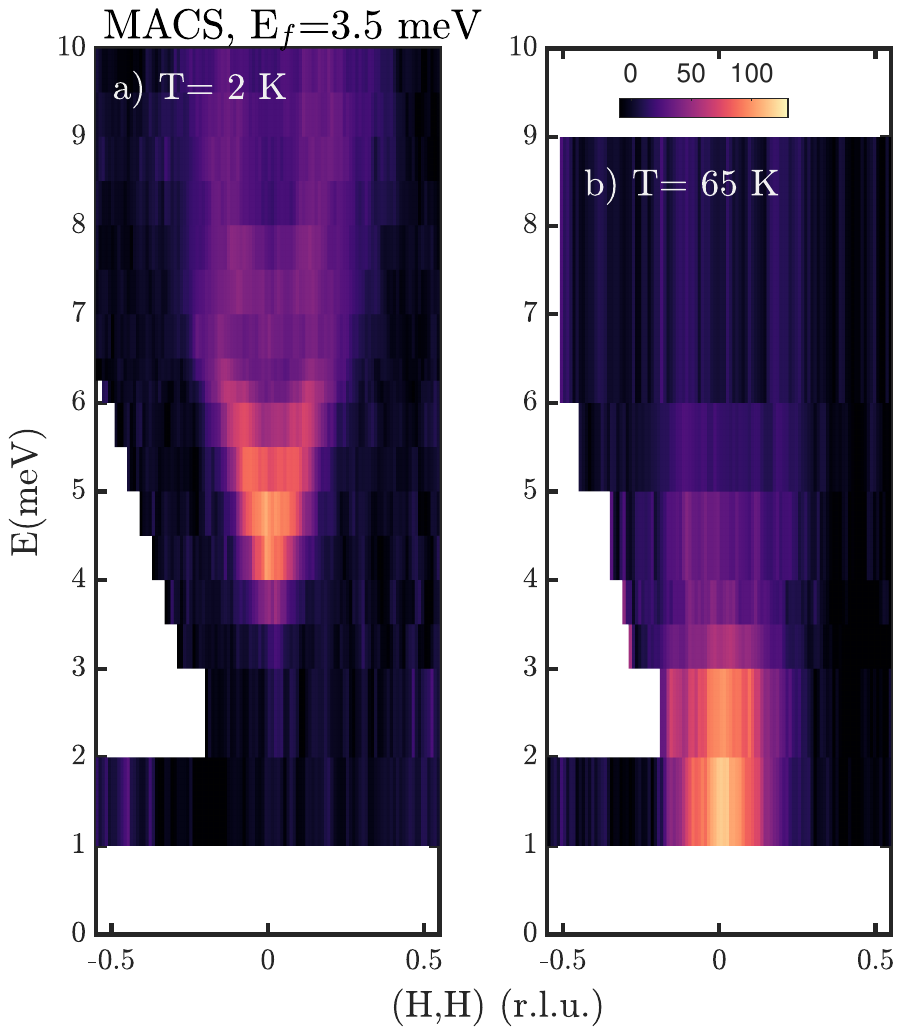}
	\caption{Ferromagnetic excitations measured on MACS and reported in Ref. \onlinecite{Lane21:104_2} in the $(a)$ magnetically ordered state at T=2 K and in the $(b)$ paramagnetic phase at 65 K.}
	\label{fig:neutron_inelastic}
\end{figure}

We first review the published neutron spectroscopy data and connect it with diffraction applying a multilevel response (or termed excitonic) model below.  There have been two experimental publications~\cite{Lane21:104_2,Gu24:132} of neutron spectroscopy in VI$_{3}$ reporting consistent results.  The low-energy magnetic excitations in the ferromagnetic ground state is presented in Fig. \ref{fig:neutron_inelastic}, previously published in Ref. \onlinecite{Lane21:104_2} utilizing the MACS neutron spectrometer at NIST.  Fig. \ref{fig:neutron_inelastic} $(a)$ illustrates the magnetic excitations at T=2 K, well below the transition to ferromagnetic order at T$_{C}\sim$ 50 K.  The spectrum displays a strong energetically gapped mode at $\sim$ 4 meV and a much weaker second mode at $\sim$ 7 meV. Both modes disperse upwards in energy for momentum transfers away from the zone center.  This was further reported in Ref. \onlinecite{Gu24:132}. At temperatures in the paramagnetic phase above T$_{C}$, Fig. \ref{fig:neutron_inelastic} $(b)$ illustrates excitations that are energetically gapless on the scale of the resolution of MACS.

\subsection{Total Magnetic Hamiltonian}

To model the neutron response function we consider the total magnetic Hamiltonian taken to be

\begin{equation}
	\mathcal{H}= \mathcal{H}_{\text{CF}} + \sum_{ij} \mathcal{J}(ij) {S}(i) \cdot {S}(j), \nonumber
\end{equation}

\noindent where $\mathit{\mathcal{H}}_{\text{CF}}$ encompasses all of the single-ion crystal field contributions and $\mathcal{J}(ij)$ denotes the Heisenberg exchange constant between spatially separated V$^{3+}$ sites $i$ and $j$. Expressing the Hamiltonian in the interaction representation yields 

\begin{equation}
	\mathcal{H}=H_1+H_2, \nonumber
\end{equation}

\noindent with $H_1$ and $H_2$ describing the single-ion and inter-ion interactions respectively. In this formalism, at the mean-field level, the single-ion term takes the form

\begin{equation}
	\mathcal{H}_1 = \sum_i \mathcal{H}_{\text{CF}}(i) + \sum_i {S}_z(i) \left( 2 \sum_{j} \mathcal{J}(ij) \langle \mathit{S}_z \rangle \right), \nonumber
\end{equation}

\noindent while the inter-ion term will be

\begin{equation}
	H_2 = \sum_{ij} \mathcal{J}(ij)[({S}_z(i)-\langle S_{z} \rangle) ({S}_z(j)-\langle S_z \rangle)+S_+(i) S_-(j)], \nonumber
\end{equation}

\noindent where $\langle S_z \rangle$ denotes a thermal averaged statically ordered spin. The above decomposition allows for the definition of a time reversal symmetry breaking molecular field 

\begin{equation}
	\mathcal{H}_{MF}(i)=\sum_i {S}_z(i) h_{MF},\nonumber
\end{equation}

\noindent with

\begin{equation} \label{mf}
	h_{MF}=2 \sum_{j} \mathcal{J}(ij) \langle \mathit{S}_z \rangle, \nonumber
\end{equation}

\noindent representing a molecular Zeeman field parameter contributing to single-ion anisotropy in the ferromagnetic phase.

\subsection{Single-Ion Hamiltonian}

Similar to previous work on VI$_3$ \cite{Lane21:104_2}, the single-ion Hamiltonian is decomposed as

\begin{equation}
	H_1= \underbrace{\mathcal{H}_{CEF}+\mathcal{H}_{SO}+\mathcal{H}_{DIS}}_{\mathcal{H}_{CF}}+\mathcal{H}_{MF}, \nonumber
\end{equation}

\noindent where $\mathcal{H}_{CEF}$ represents the contribution from the crystal electric field, $\mathcal{H}_{SO}$ from spin-orbit coupling and $\mathcal{H}_{DIS}$ from lattice distortion away from a perfect octahedral environment. $\mathcal{H}_{MF}$ was discussed above and is the molecular field contribution resulting from static magnetic order.  We now discuss each one of the contributions to $\mathcal{H}_{CF}$ which are schematically illustrated in Fig. \ref{fig:E-level splitting} $(c)$.

\subsubsection{The Crystalline Electric Field} 

In VI$_{3}$, the V$^{3+}$ ions with $L=3$ and $S=1$ are octahedrally coordinated with six I$^-$ ions, forming a $^3F$ orbital ground state. We proceed to treat the ground state manifold as a triplet with effective angular momentum $l=1$, with a projection factor from the $L=3$ manifold, $\alpha=-3/2$ \cite{Moffitt59:2}. The next excited state above the orbital ground state is at $\sim$ 1.8 eV \cite{Tanabe54:9:53,Tanabe54:9:66}, the large magnitude of the splitting ensures that mixing between the two orbital states brought upon by other contributions to the single-ion Hamiltonian will be negligible. Thus, we proceed by taking further terms in the Hamiltonian as perturbations on this $|l=1, S=1\rangle$ ground state.

\subsubsection{Spin-Orbit Coupling} 

The second key contribution to the single-ion Hamiltonian is spin-orbit coupling, which in terms of a projected angular momentum has a $|l=1,S=1\rangle$ basis, and takes the form

\begin{equation}
	\mathcal{H}_{SO}=\alpha \lambda \vec{l} \cdot \vec{S},\nonumber
\end{equation}

\noindent where $\lambda$ is the spin-orbit coupling constant and $\alpha=-{3\over 2}$~\cite{Abragam:book} is the projection factor resulting from the projection onto a $l_{eff}=1$ orbital state. The mixing between spin and orbital degrees of freedom splits the ground state into effective angular momentum values $j_{eff}=0,1,2$ (see Fig. \ref{fig:E-level splitting} $c$) as expected from the addition theorem of angular momentum. For the following calculation of the magnetic excitations, we have fixed the value from literature to $\lambda=12.9$ meV \cite{Abragam:book,McClure:book}.  While deviations from this value might be expected from covalency effects~\cite{Chakravarty59:74}, we note that this value for $\lambda$ is close to the derived value in MgV$_{2}$O$_{4}$~\cite{Lane23:5} indicating that such effects are negligible for our purposes here.

\subsubsection{Structural Distortion} 

The vanadium sites in VI$_3$ are expected to experience a distortion of the crystal electric field from the ideal octahedral coordination given the structural distortion at $\sim$ 80 K. Because of the $D_{3h}$ point group symmetry of the V$^{3+}$ ions, only the $B^0_2$ crystal field parameter is allowed in the distortion Hamiltonian \cite{Walter84:45}. Thus, it is possible to construct $\mathcal{H}_{DIS}$ in terms of Steven's operators~\cite{Hutchings64:16} as

\begin{equation}
	\mathcal{H}_{DIS}=B^0_2\mathcal{O}^0_2=\Gamma\left(l^2_z-\frac{2}{3}\right), \nonumber
\end{equation}

\noindent where the parameter $\Gamma$ characterizes the strength of the crystalline distortion away from an ideal octahedral environment.  As shown in Fig. \ref{fig:E-level splitting}, this term breaks the degeneracy of the spin-orbit levels which are further split by the molecular field present in the magnetically ordered phase.

\section{Parametrization using a Green's Function Formalism}

Having described the single-ion states and Hamiltonian, we now discuss the coupling of these spin-orbit states on the VI$_{3}$ lattice.  To parametrize the observed dispersive magnetic excitations we employ a Green's function calculation based on the random phase approximation (RPA), following closely the notation of Buyers \textit{et al.}~\cite{Buyers75:11}. Since the imaginary component of the total response function ($G^{\alpha\beta}$ with $\alpha,\beta$ being Cartesian coordinates) is proportional to the dynamical structure factor in the low temperature ($T\rightarrow 0$) limit, 

\begin{equation}
	S^{\alpha\beta}(\mathbf{q}, \omega) \propto - f^{2}(q) \mathfrak{Im}G^{\alpha\beta}(\mathbf{q}, \omega),\nonumber
\end{equation}

\noindent where $\omega$ corresponds to the energy transfer and $f(q)$ is the magnetic form factor. The Green's function gives the observed neutron scattering spectrum. This approach facilitates the multilevel nature of the spin-orbital magnetism in V$^{3+}$ and anisotropy of the crystalline electric field. A comparison of this methodology with conventional linear spin wave theory has discussed in Refs. \onlinecite{Brener24:110,Lane21:104}.  This RPA method yields the response function through a Dyson equation of the matrix form

\begin{equation}
	\underline{\underline{G}}(\mathbf{q}, \omega) = \underline{\underline{g}}(\omega) + \underline{\underline{g}}(\omega)\underline{\underline{J}}(\mathbf{q})\underline{\underline{G}}(\mathbf{q}, \omega). \nonumber
\end{equation}

\noindent where $\underline{\underline{J}}$ is a diagonal matrix proportional to the identity with elements $\mathcal{J}(\mathbf{Q})$ defined as

\begin{equation}
	\mathcal{J}(\mathbf{Q})=\frac{1}{N}\sum_{ij}\mathcal{J}(ij)\exp(-i \mathbf{Q}.(\mathbf{r}_i-\mathbf{r}_j)), \nonumber
\end{equation}

\noindent with $N$ is the number of neighboring ions. The single ion susceptibility $\underline{\underline{g}}(\omega)$ is obtained by solving the inter-level susceptibility for $\mathcal{J}(ij)=0$ to yield in the $T\rightarrow 0$ limit

\begin{equation}
	g^{\alpha\beta}(\omega) = \sum_{n} \frac{\langle 0 | S^{\alpha} | n \rangle \langle n | S^{\beta} | 0 \rangle}{\omega + i\delta - (\omega_n - \omega_0)} - \frac{\langle n | S^{\alpha} | 0 \rangle \langle 0 | S^{\beta} | n \rangle}{\omega + i\delta + (\omega_n - \omega_0)} \nonumber
\end{equation}

\noindent where the additional parameter $\delta$ moves the singularities off the real axis. Heurtistically, this parameter acts as a finite resolution for the model. 

\begin{figure}
	\centering
	\includegraphics[width=1\linewidth]{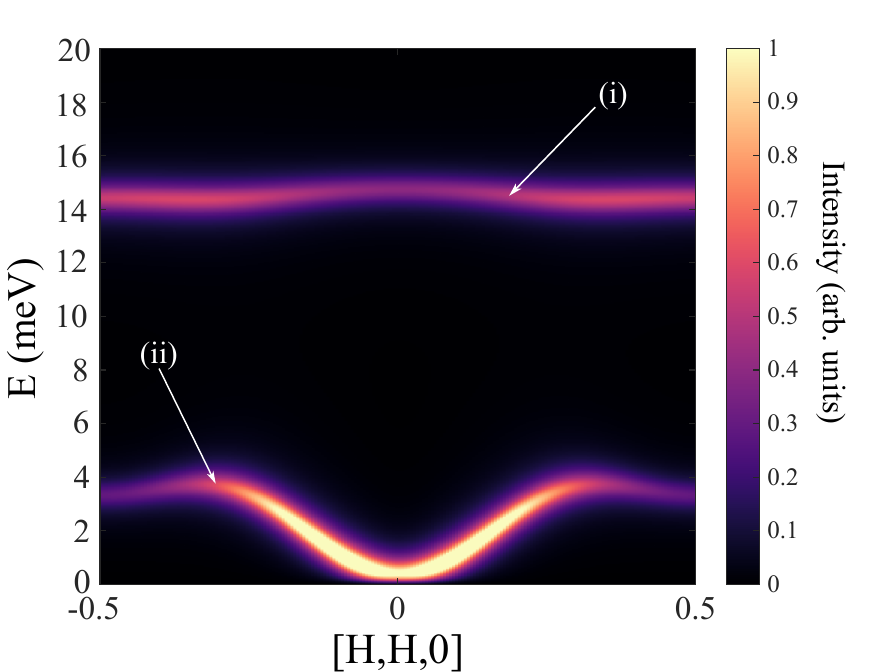}
	\caption{The magnetic dispersion relation with dominant nearest neighbour exchange displaying both a (i) optical and (ii) acoustic mode.}
	\label{fig:Optical_Acoustic}
\end{figure}

For VI$_3$, the V$^{3+}$ ions reside, at least approximately, on a honeycomb lattice which can be described as a hexagonal triangular lattice with a basis of two atoms per unit cell.~\cite{Neto09:81} The structural distortion discussed above allows for the presence of two distinct V$^{3+}$ crystalline electric field environments in the low temperature structure.  In our previous analysis discussed in Ref. \onlinecite{Lane21:104_2} we considered the superposition of two independent honeycomb lattices with differing single-ion sites to parameterize the results and nearest neighbor exchange within each domain.  This analysis relied on the upper mode being at higher energies and not observable in experiments where our current analysis does not make any assumption and also aims to be consistent with the current crystallographic data and to treat the data within a common structural framework and not invoking domains.
In the high temperature $R\overline{3}$ space group, the two V$^{3+}$ sites are related by symmetry and therefore have the same single-ion and orbital properties.  At T$_{S}$ however, this symmetry is broken and the two sites are crystallographically distinct.  Here we consider a coupled two site model on a honeycomb lattice, but with differing local single-ion environments with unequal distortion parameters $\Gamma_1$ and $\Gamma_2$ resulting from the triclinic distortion breaking the symmetry between the two V$^{3+}$ sites on the honeycomb lattice.  The idea of two distinct orbital sites is further supported by first principle calculations.~\cite{Huang20:22}  The differing local single-ion physics for the two V$^{3+}$ sites results in two energetically gapped magnetic excitations at the zone center presented in Fig. \ref{fig:neutron_inelastic} $(a)$.  Given the distinguishability of the two V$^{3+}$ sites, dominant nearest neighbor interactions would produce both an acoustic and optical mode (see Fig. \ref{fig:Optical_Acoustic}) consistent with reports in other honeycomb lattices such as CrI$_{3}$~\cite{Chen18:8}. But, such a situation is not observed in the neutron spectroscopy in VI$_{3}$~\cite{Lane21:104_2,Gu24:132} displayed in Fig. \ref{fig:neutron_inelastic} which we discuss below.

\begin{figure}
	\centering
	\includegraphics[width=1\linewidth]{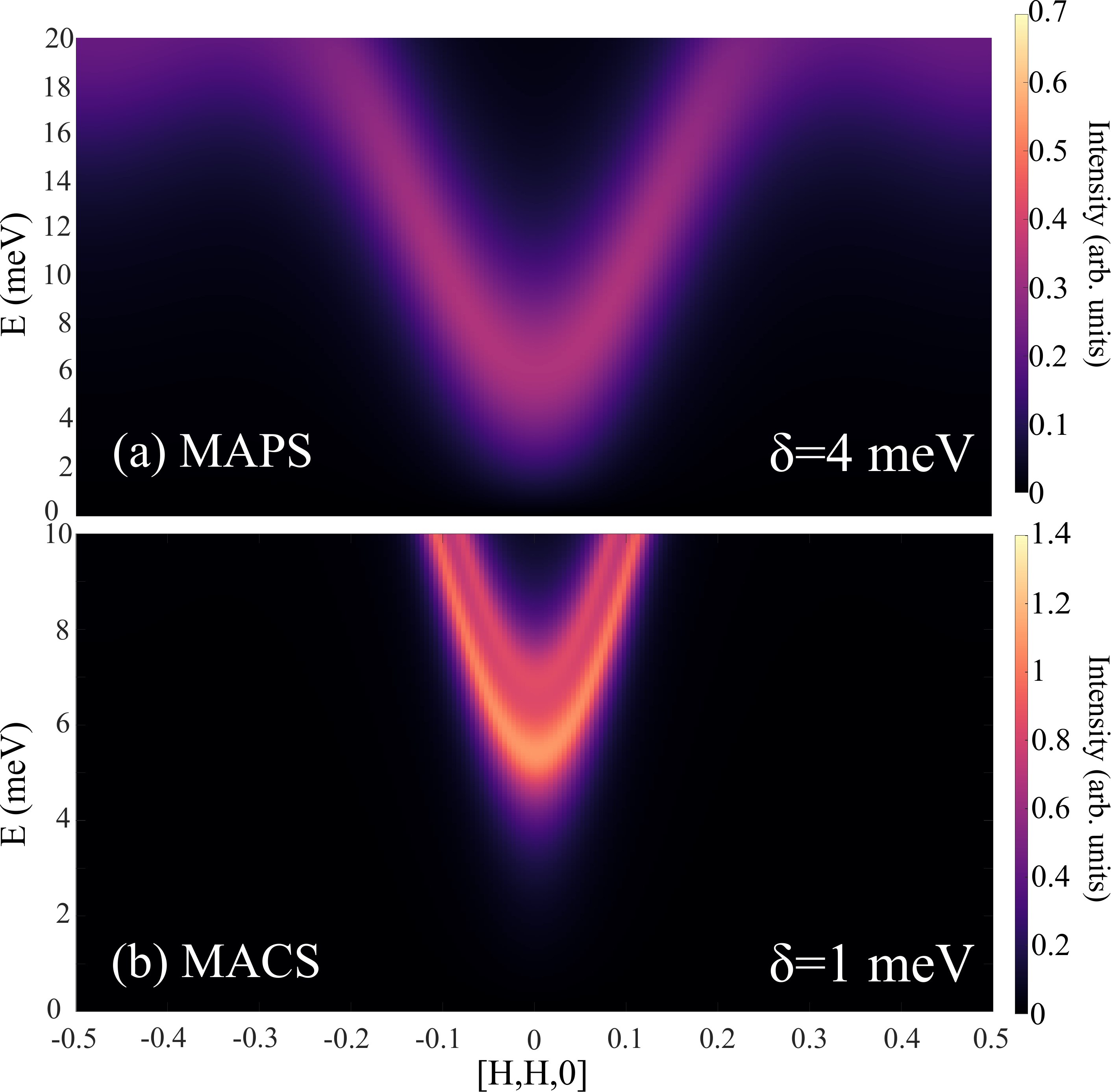}
	\caption{RPA model calculations for the data in Ref. \onlinecite{Lane21:104_2}.  The theoretical dispersion relations for the data obtained on $(a)$ MAPS~\cite{Ewings} and $(b)$ MACS spectrometers computed using the Green's function approach.}
	\label{fig:Magnetic_Theory}
\end{figure}

\begin{figure*}[!htbp]
	\centering
	\includegraphics[width=140mm,trim=2.5cm 4.0cm 2.5cm 3.5cm,clip=true]{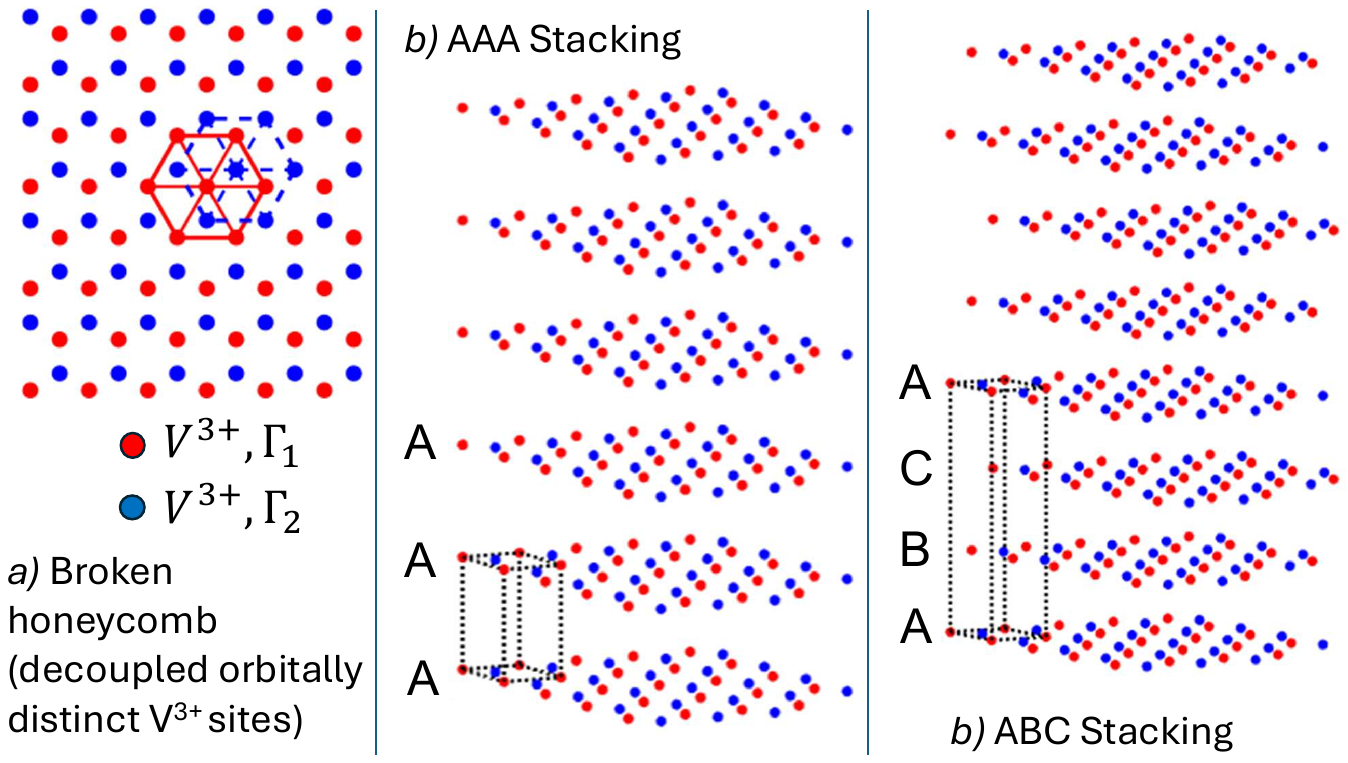}
	\caption{$(a)$ Coordination of the broken honeycomb lattice in VI$_{3}$.  The structural breaking breaks up the coordination on the V$^{3+}$ sites from being threefold coordinated (like in a honeycomb) to sixfold (like in a triangular arrangement).  $(b)$ $AAA$ stacking schematic like in VI$_{2}$.  $(c)$ $ABC$ stacking schematic present in VI$_{3}$.  We speculate that the combination of symmetry breaking of the honeycomb lattice and changes in stacking may create a situation where VI$_{3}$ behaves more like VI$_{2}$ and displays an additional $\sim$ 30 K transition.}
	\label{fig:Broken_honeycomb}
\end{figure*}

However, considering dominant next-nearest neighbor exchange where only crystallographically identical V$^{3+}$ sites are coupled with only very weak nearest neighbor exchange, it is possible to reproduce the observed dispersion relations (see Fig. \ref{fig:Magnetic_Theory}). The selective exchange between sites is best explained through the super-exchange interaction, similar to what has been observed in some oxides \cite{Anderson50:79,Goodenough55:98,Kramers34:1}, where the maximal overlap between the $p$ and $d$ orbitals of the anion and cation, respectively, are found for the 180$^\circ$ in-plane V-I-I-V orientation observed for next-next-nearest neighbor interactions. This is in contrast to the minimal overlap observed for the $\sim 90$$^\circ$ V-I-V arrangement for nearest neighbors (see Fig. \ref{fig:E-level splitting} $(a)$). 

Fixing the distortion parameters to $\Gamma_1=-28$ meV and  $\Gamma_2=-20$ meV (see Figs. \ref{fig:Magnetic_Theory} $a,b$). The total set of parameters obtained from the RPA model are presented in Table \ref{Table_Parameters}.  The parameters show that the distortion parameter dominating over the spin-orbit contribution ($\mathcal{H}_{DIS} > \mathcal{H}_{SO}$) in the single-ion Hamiltonian.  This is consistent with a structural distortion occuring at higher temperatures independent of magnetic ordering.   We note that a similar situation has been proposed for the magnetic and structural transitions in MgV$_{2}$O$_{4}$.~\cite{Lane23:5}   In the case where ($\mathcal{H}_{DIS} < \mathcal{H}_{SO}$) we would expect magnetic and structural phase transitions to occur at the same temperature as observed in, for example, CoO~\cite{Tomiyasu04:70}.  The large distortion terms here imply large anisotropy (reflected in the magnetic dynamics) and is consistent with suggestion that VI$_{3}$ is close to a three- to two-dimensional critical point.~\cite{Liu20:1}

The dominant next nearest exchange model which couples V$^{3+}$ sites with the \textit{same} single-ion environments (hence same distortion parameter $\Gamma$) reproduces the magnetic dispersion near the zone center.  Of particular note is the inclusion of a small $\mathcal{J}_1$ results in an intensity mismatch between the two modes which agrees with experiment.  We discuss this in connection to the diffraction results below.

While Gu \textit{et al.}~\cite{Gu24:132} interpret the neutron spectra of VI$_3$  within a uniform honeycomb framework requiring a multi-parameter Hamiltonian ($J$--$K$--$\Gamma$--$\Gamma'$--$A$, termed the PRL model here), our analysis emphasizes the structural distortion at $T_{s} \approx 80$~K that produces two crystallographically 
inequivalent V$^{3+}$ sites. In the PRL approach, the complexity of the excitation 
spectrum is captured by introducing several symmetry-allowed anisotropic exchanges,  effectively treating the two observed branches as arising from distinct orbital manifolds but within separate fitted models. By contrast, our magnetoelastic 
fragmentation picture naturally accounts for the presence of two modes through 
inequivalent local single-ion environments, each with its own spin-orbital ground 
state. In this framework, the dominant physics is encoded in the single-ion 
splitting and exchange between symmetry-equivalent sites, reducing the need for 
multiple adjustable exchange parameters. Thus, the two descriptions differ in 
emphasis: the PRL model distributes the complexity into an exchange Hamiltonian 
on a uniform lattice, whereas our model locates it in the orbital-driven 
fragmentation of the honeycomb.

\begin{table}[h] \label{Table_Parameters}
	\begin{tabular}{@{}l@{\hspace{20pt}}l@{}}
		\multicolumn{2}{c}{Green's Function Model Parameters} \\ 
		\hline
		Parameter & Value \\ 
		\hline 
		$\mathcal{J}_1$ & -0.1$\pm0.04$ meV\\ 
		$\mathcal{J}_2$ & -2.5$\pm0.2$ meV\\
		$h_{MF}$ & 12$\pm1.5$ meV\\
		$\lambda$ & 12.9 meV~\cite{Abragam:book} \\
		$\alpha$ & -3/2\\
		$\Gamma_1$& -28$\pm1$ meV\\
		$\Gamma_2$& -20$\pm1$ meV\\
		\hline
	\end{tabular}
	\caption{Table displaying the parameters together with their respective uncertainties obtained from the Green's function model.}
	\label{Table_Parameters}
\end{table}

\section{Discussion}

We have argued for the presence of a single structural transition in VI$_{3}$ based on a comparison between powder samples made under different conditions and single crystal samples synthesized by vapor transport.  Through Le Bail fits to powder data combined with group theory, we find this transition characterizes a change in the space group from a rhombohedral ($R\overline{3}$) unit cell to triclinic (either $P1$ or $P\overline{1}$).  This structural transition is followed by a lower temperature ferromagnetic transition which displays measurable magnetoelastic coupling.  This taken with the structural transition is used to argue that spin-orbit coupling is a relevant term in the magnetic Hamiltonian and provided the foundation for our multilevel Green's response function analysis of the dynamics presented above.  

At high temperatures in the $R\overline{3}$ phase, the two V$^{3+}$ are symmetry related with the same crystalline electric field environment and single-ion Hamiltonian.  Below T$_{S}$, this symmetry is broken on entering the triclinic phase resulting in two crystallographically distinct sites bringing about a broken honeycomb lattice at low temperature.   This is further validated by our analysis of the magnetic dynamics which provides evidence for two distinct V$^{3+}$ sites.  Notably, we observe that crystallographically distinct sites are magnetically coupled to the same single-ion site which effectively fragments the honeycomb lattice into two interpenetrating hexagonal lattices with a nearly triangular motif.  In terms of the V$^{3+}$ coordination, this alters the coordination from being threefold (in the high temperature honeycomb phase) to being sixfold (in the low temperature triclinic phase).  This situation is schematically illustrated in Fig. \ref{fig:Broken_honeycomb} $(a)$ which provides an illustration of the two distinct V$^{3+}$ sites present at low temperatures and the dominant coupling found with neutron spectroscopy.

While the two-site model is compelling based on our structural data, group theoretical analysis, and spectroscopy, there remain points that warrant caution. Our conjecture of two sites is supported by Le-Bail fits to the unit cell shape combined with neutron spectroscopy, although preferred orientation effects limited the reliability of full Rietveld refinements of atomic positions. Neutron spectroscopy on single-crystal samples that display only one structural transition nevertheless provides strong evidence for two distinct sites with differing single-ion properties. NMR data in Ref. \onlinecite{Gati19:100} report two sites below $\sim$36 K but only a single site above this temperature. While this has been interpreted as evidence for a lower structural transition, we suggest that the higher-temperature behavior is not consistent with structural studies and may instead reflect evolving single-ion properties, as illustrated in Fig.  \ref{fig:E-level splitting}.

The breaking of the symmetry between the two V$^{3+}$ ion sites resulting in distinct local single-ion environments results in two distinct orbital sites.  As discussed in Ref. \onlinecite{Solovyev24:110} this has direct consequences on the macroscopic properties with ferroelectricity being suggested.  Given our current powder data we are not able to distinguish between $P1$ (ferroelectric without an inversion center) and $P\overline{1}$ (with an inversion symmetry as tabulated in Table \ref{Table:symmetry_elements} and hence not ferroelectric).  Future single crystal experiments will be required to make this distinction.  

We now use our analysis of the spin excitations combined with our diffraction results to speculate as to the origin of the T$\sim$ 30 K transition found in some of our powder samples and reported in the literature.  In our samples, the materials that displayed the most prominent evidence of this transition also had additional Bragg peaks that could be indexed to large Miller index $L$ values.  This could indicate a different stacking or more ordered stacking arrangement than in our other samples. It might also be consistent with the short-range order reported in VI$_{3}$.~\cite{Mijin20:59}  Given that VI$_{2}$~\cite{Kuindersma79:30} and related compounds show magnetostructural transitions over the same temperature, we propose that the combination of the breaking up of the honeycomb into two effectively interpenetrating hexagonal lattices and more perfect stacking of van der Waals layers, may provide a situation where VI$_{3}$ mimics its hexagonal VI$_{2}$ counterpart.   While our results are based on diffraction and spectroscopy, we note that layer dependent magnetism has been noted in CrI$_{3}$ applying the Kerr effect.~\cite{Huang17:546} Direct imaging using real-space probes would be highly desirable to confirm these conclusions. We emphasize that this is highly speculative, but stacking dependent properties and phase transitions have been proposed recently in CrI$_{2}$~\cite{Zhang22:105}.

In summary, we have investigated the magnetoelastic symmetry breaking in VI$_{3}$.  We observe the breaking of the $R\overline{3}$ at T$_{s} \sim$ 80 K and a lower ferromagnetic transition at T$_{C} \sim$ 50 K.  While we observe a further transition at $\sim$ 30 K, it is not reproduceable in our samples and also we propose it is inconsistent with group theoretical arguments based on subgroups.  Through an investigation of the dynamics, we find the structural transition breaks the honeycomb lattice up into two interpenetrating hexagonal lattices with crystallographically distinct orbitally active V$^{3+}$ sites which we refer to as a fragmented honeycomb. 

\section{Acknowledgements}

We acknowledge funding from the EPSRC, Royal Society of Edinburgh, and the STFC.  

\section{Data Availability}

The data that support the findings of this study are available from the corresponding author upon reasonable request.~\cite{datarequest}


%

\end{document}